\documentclass[fleqn, usenatbib]{mnras}
\usepackage{flushend}
\usepackage{newtxtext,newtxmath}
\usepackage{soul}
\usepackage{hyperref}
\usepackage{float}
\usepackage{mathtools}
\usepackage{lipsum}
\usepackage{ulem} 
\usepackage{graphicx}
\usepackage{subfig, caption}
\usepackage[T1]{fontenc}
\usepackage{ae,aecompl}
\usepackage{amsmath}
\usepackage{amssymb}
\usepackage{scalerel}
\usepackage{tikz}
\usetikzlibrary{svg.path}
\definecolor{orcidlogocol}{HTML}{A6CE39}
\tikzset{
	orcidlogo/.pic={
		\fill[orcidlogocol] svg{M256,128c0,70.7-57.3,128-128,128C57.3,256,0,198.7,0,128C0,57.3,57.3,0,128,0C198.7,0,256,57.3,256,128z};
		\fill[white] svg{M86.3,186.2H70.9V79.1h15.4v48.4V186.2z}
		svg{M108.9,79.1h41.6c39.6,0,57,28.3,57,53.6c0,27.5-21.5,53.6-56.8,53.6h-41.8V79.1z M124.3,172.4h24.5c34.9,0,42.9-26.5,42.9-39.7c0-21.5-13.7-39.7-43.7-39.7h-23.7V172.4z}
		svg{M88.7,56.8c0,5.5-4.5,10.1-10.1,10.1c-5.6,0-10.1-4.6-10.1-10.1c0-5.6,4.5-10.1,10.1-10.1C84.2,46.7,88.7,51.3,88.7,56.8z};
	}
}

\newcommand\orcidicon[1]{\href{https://orcid.org/#1}{\mbox{\scalerel*{
				\begin{tikzpicture}[yscale=-1,transform shape]
					\pic{orcidlogo};
				\end{tikzpicture}
			}{|}}}}
\newcommand\scalemath[2]{\scalebox{#1}{\mbox{\ensuremath{\displaystyle #2}}}}
\title[DM decay/annihilation and tSZ effect ]{Thermal SZ effect in a magnetized IGM dominated by interacting DM decay/annihilation during dark ages}
\author[Pandey \& Malik 2022]{
	Arun Kumar Pandey \orcidicon{0000-0002-1334-043X}$^1$
	\thanks{E-mail: arunp77@gmail.com, (arun\_pandey@prl.iitgn.ac.in)},
	Sunil Malik \orcidicon{0000-0003-4147-626X} $^{2,3}$
	\thanks{E-mail: sunilmalik@uni-postdam.de}
	\\
	$^{1}$Department of Physics and Astrophysics, University of Delhi, New Delhi-110 007, India,\\
	$^{2}$ 	Institute fur Physik und Astronomie	Universitat Potsdam, Haus 28, Karl-Liebknecht-Str 24/25, D-14476 Potsdam, Germany,\\
 $^3$Deutsches Elektronen-Synchrotron DESY, Platanenallee 6, 15738 Zeuthen, Germany.
}
\date{\today}
\begin{document}
	\maketitle
\begin{abstract}
During cosmic dawn, the thermal history of the universe is well studied, and a study of this era can give us some of the most useful insight into the universe before the recombination epoch. Its precise modeling and future high-precision measurements will be a valuable tool for determining the thermal history of the universe. In the present work, we study the thermal and ionization history of IGM in the presence of decaying magnetic fields via ambipolar and turbulent decay, Baryon-Dark matter (BDM) interaction, including the DM decay/annihilation. The BDM interaction cross-sections considered are of the form $\sigma=\sigma_0 v^{n}$, where $n=-2$ and $n=-4$. In this work, we show that in the current scenario, the decay/annihilation of the DM particles have a considerable impact on the temperature and ionization histories at low redshift. With the addition of the concept of fractional interaction, which states that if a fraction of the DM particles interacts with the baryons, the temperature and ionization fraction of the baryons show a strong dependence on the percentage of DM particles interacting with the baryons. We have also studied the interesting consequences of the present scenario on the thermal Sunyaev-Zeldovich (tSZ) effect. We show that the highest value of the absolute value of the mean $y-$parameter in the current DM decay/annihilation scenario is well within the values derived from experimental data such as PLANCK, FIRAS, and PIXIE. Later we calculate the bound on the ordinary magnetic fields originating from the Dark photons.
\end{abstract}
\begin{keywords}
	SZ effect, CMB, dark matter, IGM, magnetic fields, ambipolar diffusion, turbulent decay, dark matter decay/annihilation, baryon-DM interaction, Dark photons
\end{keywords}
\section{Introduction}\label{sec:intro}
In the standard $\Lambda$CDM model of cosmology, thermal and the ionization history of the universe, between the recombination epoch and the first-star formation, is well studied \cite{Peebles:1968ja, Zeldovich:1969en, AliHaimoud:2010dx, Chluba:2011rm}. After the recombination epoch, due to Compton scattering between the residual free electrons and the CMB photons, temperature of the matter $T_m$ and CMB temperature $T_{\rm cmb}$ remains same, i.e. $T_m \approx T_{\rm cmb}$ till redshift $z\sim 100$. Subsequently, the interaction rate becomes so small that matter and CMB photons decouple, and the temperature of the matter evolves with redshift as $T_m(z)\propto (1+z)^2$. This behavior of matter temperature and hence the ionization fraction can change significantly and can evolve differently if there is some other factor involved. Some important factors are magnetic fields, dark matter (DM)-baryon interactions, and DM decay or annihilation in the IGM. Any deviation from the standard history is thus an indication of a new mechanism of heating and cooling. The `Experiment to Detect the Global Epoch of reionization Signature' (EDGES) collaboration has discovered an absorption trough signal at $78$ MHz in the $14< z< 20$ shift region in a 21-cm signal \cite{Bowman:2018yin}. The measured trough in the 21-cm absorption at $z\sim 17.2$ is $T_{21} = -500^{+200}_{-500}$ mK with 90\% confidence level. This observation of signals by the collaborations opens a new window into the dark ages, shedding new light on the thermal and ionization history. Since the publication of the EDGES results, there were several models developed, considering colder than expected gas temperature or additional sources of photons. There were few models given where baryons and dark matter (represented by $\chi$ in the upcoming sections) interactions are considered to lower the gas temperature by cooling it by absorption of heat from the gas to DM (assuming DM to be colder than gas) \cite{Munoz:2015bk, Munoz:2017qpy, Barkana:2018nd, Pandey:2020hfc}. The most commonly studied portal through which the interaction happens is via some mediator, for example, a vector mediator. In this case, interaction takes place because of the kinetic mixing between one dark and one ordinary Abelian gauge boson. The ordinary gauge bosons are considered to be from the $U(1)$ gauge group of the standard model or hypercharge fields above the electroweak symmetry breaking scale. However, the photon of the dark sector comes to be identified as the boson of $U(1)^\prime$ group of the dark sector. The coupling between the standard model particles (SMP) and Dark matter particles (DMP) may occur through following channels: $\text{SMP}~ \xrightarrow{e}~A_\mu \xrightarrow{\epsilon e'}~ \text{DMP}\xrightarrow{e'} A_\mu'$ or  $A_\mu \xrightarrow{e} ~\text{SMP} \xrightarrow{\epsilon e} ~A_\mu'~\xrightarrow{e'} ~ \text{DMP}$ (where $e$ and $e'$ are the coupling constant s of the ordinary and dark photons respectively sectors and $\epsilon=\text{sin}\theta$, $\theta$ being the rotation angle). The low energy Lagrangian of this interaction could be written as \cite{Chen:2020bok}
%
\begin{eqnarray}
	\mathcal{L} & = & \mathcal{L}_{\rm SM}+\mathcal{L}_{\rm DM}-\frac{1}{4}X_{\mu\nu}\, X^{\mu\nu}+\frac{1}{2} m^2_\chi X_\mu X^\mu \nonumber \\
	& - & \frac{\epsilon}{2}F_{\mu\nu} X^{\mu\nu}+ \epsilon e J_\mu X^\mu +\epsilon e' J'_\mu A^\mu
\end{eqnarray} 
where $\mathcal{L}_{\rm SM}$ and $\mathcal{L}_{\rm DM}$ are the Lagrangian belonging to the SM particles and the DM particles respectively. The third and fourth terms are the kinetic and mass terms for the dark photons, respectively. The fifth term is nothing but a kinetic mixing of dark photon field $A_\mu'$ and the ordinary photon field $A_\mu$ \cite{Masaki:2018eut, Berezhiani:2013dea, Kandus:2010nw}. The last two terms are the coupling between SM  and dark photons and vice versa. The product of two gauge groups $U(1)\times U(1)'$ is eventually broken down to some diagonal $U(1)_{\rm em}$ group which represents the ordinary massless photon eigenstate $\gamma$ while another mass eigenstate $\gamma'$ becomes heavy. This ordinary photon may interact primarily with SM particles with the coupling constant $e$, but it will perceive DM particles as having electric charges $e'=\epsilon e$ \cite{Holdom:1985ag}.

To cool the gas, a Rutherford-like interaction cross-section between the DM and the gas particles is considered \cite{Prinz:1998ba, Spergel:2000nj, Davidson:2000hf, Dubovsky:2004ds, Tulin:2013hb, Barkana:2018lgd}. A generalized form of the  interaction cross-section is given by $\sigma (v)=\sigma_0(v/c)^n$ (where $v$ is the relative velocity of the DM and gas particles), $n=-1$ for the Yukawa potential (massive Boson exchange) \cite{Aviles:2011ac}, $n=-2$ is for an electric dipole moment \cite{Sigurdson:2004kd} and $n=-4$ corresponds to millicharge DM particles \cite{McDermott:2011dy, Dolgov:2013ad}. Above velocity-dependent DM and gas particles interaction seems a best way to evade present day, astrophysical and cosmological constraints \cite{Jaeckel:2010ni}, including the constraint on relativistic degree of freedom from the  big bang nucleosynthesis (BBN) and CMB \cite{Davidson:2000hf, Vogel:2014jv}, anisotropy in CMB \cite{Dubovsky:2004ds, Dolgov:2013ad, Dvorkin:2013cea}, fifth force experiments \cite{Adelberger:2003zx, Williams:2004qba, Kapner:2006si, Schlamminger:2007ht,  Xu:2018efh}, the stellar cooling \cite{Hardy:2016kme} and bounds on the self-interactions of DM \cite{Tulin:2013yh, Tulin:2013hb}. Based on a possible spectral distortion of CMB photons, due to the interaction of the dark matter with the baryons, bounds on the spectral distortion are given for the interaction cross-section \cite{McDonald:2000tp}. Self-interactions and the interaction with the standard model (SM) particles are the deciding factors for the velocity distribution of the DM. In the $\Lambda$CDM universe, the cold DM and baryon fluids have a relative velocity, resulting in supersonic coherent baryon flows after recombination \cite{Tseliakhovich:2010dh}. If the momentum transfer rate between DM and the baryons while interacting is low, the drag between the two fluids may not efficiently dissipate their relative bulk velocity. The relative motion dominates over the thermal motions when the universe is sufficiently cooled down. Furthermore, if the relative bulk velocity is considerably large before recombination, the Boltzmann equations used to understand the fluid velocity fluctuations become nonlinear due to mixing the individual Fourier modes. To avoid the mixing problem for our convenience, we have used the root mean square velocity of the relative bulk velocity as a correction to the thermal velocity dispersion \cite{Dvorkin:2013cea, Xu:2018efh, Slatyer:2018aqg}. The frequent self-interaction causes DM velocities to reshuffle towards the thermal Maxwell-Boltzmann (MB) distribution, with the highest entropy. The DM velocity distribution does not have to be MB when the DM thermally decouples from SM particles if the timescale for velocity redistribution through self-interactions is longer than the expansion period \cite{Yacine::2021amy}. For a limiting case when the self-interactions are negligible, the heat exchange between the DM particles with baryons can differ by up to a factor of $\sim 2-3$ only \cite{Yacine:2019alh}. 

It is known that magnetic fields of different amplitudes are present at all length scales \cite{Pandey:2015kaa, Subramanian:2015lua, Anand:2017zpg, Pandey:2021lru}. The origin of the seed magnetic field remains an open-ended subject in current cosmology, despite having so many fundamental impacts. The pervasive magnetic fields are anticipated to have a significant impact on a variety of processes, including baryogenesis \cite{Giovannini:1998mem, Simone:2011ads, Kushwaha:2021aks}, primordial nucleosynthesis \cite{Grasso:1996kk}, structure formation \cite{Sethi:2004pe, Kahniashvili:2012dy, Katz:2021iou}, and cosmic microwave background physics \cite{Barrow:1997mj, Yamazaki:2010dsy, Kunze:2014eka} (for more details, see the review article \cite{Grasso:2000wj}). In reference \cite{Berezhiani:2013dea}, authors have discussed the origin of these fields during galaxy formation, when a finite interaction between the baryons and DM particles is considered. The mechanism discussed here has some implications for our present work in the context of DM particles. The amplitude and the properties of these magnetic fields depend on the origin mechanism, and it is characterized by present-day field strength $B_\lambda$ at a given length scale $\lambda$. If these magnetic fields are present at the time of decoupling, due to the Lorentz force in the ionized plasma, significant density fluctuations can be generated by Magnetohydrodynamic motions and heat the gas. Heating of the gas occurs from the ambipolar diffusion and turbulent decay of the magnetic fields \cite{Sethi:2004pe}. The heating of gas affects the structure formation and hence can affect the reionization and other further processes after the decoupling epoch.  \cite{Cowling:1956gt, Wasserman:1978iw, Kim:1994zh, Sethi:2004pe}. The effect of these magnetic fields on the neutral hydrogen in the IGM by hydrodynamic effects is used to constrain the strength of the magnetic fields \cite{Tashiro:2005ua}. 

In addition to energy deposition from the interaction of the DM and gas particles, DM decays or annihilation may also inject energy \cite{Mapelli:2006ej, Galli:2009zc, Giesen:2012rp, Lopez-Honorez:2013cua, Ade:2015xua, Slatyer:2016rj,  Liu:2018uzy, Fraser:2018acy, Mitridate:2018iag, Aghanim:2018eyx}. Annihilation/decay of the DM could inject electromagnetically interacting particles into the universe, with a potentially wide range of observable consequences. These ionizing particles during the cosmic dark ages may change the ionization and the thermal history along with the last scattering surface width and hence can modify the anisotropy of the cosmic microwave background (CMB) \cite{Adams:1998jw, Chen:2003gz, Padmanabhan:2005pf}. One of the proposed origin mechanisms of magnetic fields, as stated above, is when DM particles interact with baryons in some way. The extra energy induced due to the DM annihilation or decay will leave a noticeable impact on the 21-cm signal and its fluctuations \cite{Short:2019twc}. Recently,  it has been shown that energy injection by the DM annihilation impacted CMB, and a novel way of constraining DM annihilation, based on PLANCK 2018, is discussed \cite{Huang:2021voi}. The discussion thus far has been focused mainly on the single species of DM, weakly interacting with the baryons at all times. However, it is known that the bounds on interaction cross-section from the CMB suggest weak coupling at all times only for certain species of the DM particle (i.e., specific values of `\textit{n}'). It is known that for redshift $z \gtrsim 10^4-10^5$, coupling is stronger for $n\geq 0$ \cite{Gluscevic:2017ywp, Boddy:2018kfv}.  In this era, the baryon and DM remain strongly coupled with each other and the DM behaves as an additional baryonic component, except that they do not participate in the recombination process. For $n=-2$ and $n=-4$, the CMB bounds on $\sigma_0$ does not violate, if we assume a fraction $f_{\chi_{,\rm in}}=\rho_d/\rho_{\rm DM}$ of the total DM density $\rho_{\rm DM}$ only interact with the baryons. In this case, the remaining fraction of the DM is cold collisionless DM \cite{Boddy:2018wzy}. In the present work, we have studied thermal and ionization history for various values of $f_{\chi_{,\rm in}}$. A value of $f_{\chi_{,\rm in}}=1$ means all the DM particles are interacting with the baryons. The baryon-DM interaction is considered under the following assumption: (i). the DM particle has no interaction with the neutral component of baryons, implying that it has little interaction with atomic dipole moments, (ii). the interactions between the charged components and the rest of the matter (and within the remaing matter) are large enough to maintain a constant velocity and temperature for all baryonic components (iii). A nonzero and tiny electric charge exists in a portion of the DM (i.e., $f_{\chi,{\rm int}}<1$) that does not interact with the neutral fraction (iv). the charged component of the DM particles is a Fermion with mass $m_\chi >1$ MeV and has a fractional charge, and the positively and negatively charged components have particle-antiparticle asymmetry.

In summary, to study the thermal and ionization history of IGM, we have considered the following effects in addition to the cooling effects from Bremsstrahlung, collisional excitation, recombination, and collisional ionization cooling:
\begin{itemize}
    \item extra energy injection from the annihilation/decay of the DM particles,
    \item decaying magnetic fields in an early universe plasma via ambipolar diffusion and turbulent processes,
    \item energy exchange between the baryons and DM particle due to their non-standard velocity-dependent interactions.
\end{itemize}
Since the thermal Sunyaev–Zeldovich (tSZ) effect directly probes the thermal energy of the universe, precise modeling and future high-accuracy measurements will provide a powerful way to constrain the thermal history of the universe. The tSZ effect occurs when ionized electrons with thermal motion scatter CMB photons, resulting in secondary CMB temperature oscillations. The evolution of the $y-$parameter, which quantifies the distortion in the CMB, with respect to redshift $z$ is presented in the last section (see subsection \eqref{sec-3-tsz-effect}). We also compared our findings to previous research on the temperature and ionization history of the universe during the dark ages. In the current study, we mainly focus on the redshift $10< z < 1100$ to avoid the non-linear evolution of the density perturbations. This work is divided in following sections: section (\ref{sec-2}), contains the thermal and ionization history, where we have address baryon-DM interaction (see subsection (\ref{sec-2-sub-2.3})) along with the DM decay/annihilation (\ref{sec-2-sub-2.4}) and heating of IGM due to the ambipolar and turbulent decay of magnetic fields (\ref{sec-2-sub-2.2}) in addition to the standard scenario (\ref{sec-2-sub-2.1}). Finally, we have discussed the obtained results in section (\ref{sec-3}) and concluded the work in section (\ref{sec-4}). We want to mention here that we are working with the MKS unit. In the present work, we have considered the flat FLRW universe and used following present-day parameters: $H_0=h\times 100$ km s$^{-1}$ (where $h=0.6766$), $h^2 \Omega_b =0.022$, $h^2 \Omega_m =0.142$, and $h^2 \Omega_\Lambda =0.318$, Boltzmann constant $k_B=1.381\times 10^{-23}$ J K$^{-1}$, and speed of light $c=3\times 10^8$ m s$^{-1}$ \cite{Aghanim:2018eyx}.
\section{Ionization and thermal history of the universe}
\label{sec-2}
There is still a major gap in our understanding of the IGM's thermal and ionization histories after the recombination epoch. Various theoretical studies are given to understand the evolution of gas in the IGM. Within the $\Lambda$CDM model of cosmology, at the end of the recombination epoch, matter decouples from the CMB photons and the universe temperature reaches around $3000$ K. After this epoch, the Universe becomes for most of the time, homogeneous until the epoch of reionization. This period is known as the dark ages. During this period of time, residual electrons leftover from the recombination scatter with the baryons and maintain the thermal equilibrium until $z\approx 200$. As a result of the Universe's thermal expansion, the gas cools adiabatically, reaching a temperature of 10 K at redshift $z\approx 17$. Later, during the epoch of reionization, over-density in the matter perturbations grows and eventually collapses to form the first star and galaxy. In the present work, we will study the implications of the DM decay or annihilation on the thermal and ionization history, in conjunction with the three other mechanisms that could deepen the 21-cm absorption signal (detected by EDGES collaboration): i). nonstandard recombination histories (with magnetic fields); ii). baryon-DM interaction; and iii). an additional source of 21-cm photons during the dark ages. These mechanisms were given to explain the EDGES measurements \cite{Bowman:2018yin}. In our present work, we include all these effects to study their role in the thermal and ionization history of the IGM. 

This section is divided into the following subsections. In subsection (\ref{sec-2-sub-2.1}), we have presented the well-known standard thermal history of the universe. In the next subsection (\ref{sec-2-sub-2.2}), we have briefly discussed baryon-DM interaction and heating of the gas due to a finite difference in the temperature of baryons and the DM particles; and a non-zero relative velocity. Subsection (\ref{sec-2-sub-2.3}) contains the heating of the gas due to magnetic heating from the ambipolar and turbulent decay. The last subsection (\ref{sec-2-sub-2.4}) of this section briefly summarizes the energy injection in the system by deposition due to the annihilation or decay of the DM particles. At the end of this section, we have written a complete set of equations required for our study in subsection (\ref{sec-2-sub-2.5}).
\subsection{Standard scenario}
\label{sec-2-sub-2.1}
During and after the recombination epoch, the conventional perception of the thermal and the ionization history of the universe is well approximated by the three-level atomic model \cite{Peebles:1968ja, Zeldovich:1969en, AliHaimoud:2010dx}. In this standard theory, matter and radiation are in thermal equilibrium before the recombination epoch ($z\gg 1100$). After this epoch, the interaction between the matter and CMB photons becomes so weak that matter decouples from the CMB photons and reaches $\sim 3000$ K. After this epoch, dark ages start, and few residue electrons and CMB photons are the main ingredients. The residual free electrons allow Compton scattering to maintain a thermal coupling of the gas to the CMB photons and set kinetic temperature of the gas $T_b$ equals to CMB temperature $T_{\rm CMB}$ between redshift $200 \lesssim z \lesssim 1100$ \cite{Pritchard:2011xb}. The gas cools adiabatically as $T_b\propto (1+z)^2$ for $z_* \lesssim z \lesssim 200$ ($z_*$ is the redshift at which first star formation takes place or additional energy injected to reionize the system) \cite{Hirata:2006bn, Loeb:2003ya}. With the expansion of the universe, the number density of the gas and collisional coupling becomes weak, and hence $T_b < T_{\rm CMB}$. Below redshift $z\lesssim z_*$, for energy injected from the first-star formation or any other source, kinetic temperature $T_b$ becomes larger than the CMB temperature. The evolution equations of the baryonic matter and the ionization fraction is given by
 \cite{AliHaimoud:2010dx}
\begin{eqnarray}
\frac{dT_{b}^{\rm sta}}{dz} & = & \frac{2T_b}{(1+z)} + \frac{\Gamma_{C}}{(1+z)H} (T_b-T_{\rm CMB}), \label{eq:baryon_temp-1}\\
\frac{dx_{e}^{\rm sta}}{dz} & = &  \frac{1}{H(1+z)} \Bigg[\left(n_{\small B} x_e^2\alpha_{B}-(1-x_e)\beta_{B}e^{-E_{21}/T_{\rm CMB}} \right) D \nonumber \\
& - & \gamma_e\, n_{\small B} (1-x_e)\, x_e \Bigg].
\label{eq:ioni1}	
\end{eqnarray}
In equation (\ref{eq:baryon_temp-1}), $\Gamma_C$ is Compton scattering rate and it is defined as \cite{Yacine:2010acm}
\begin{equation}
	\Gamma_C =\frac{8\sigma_T x_e \rho_{\rm CMB}}{3(1+x_e+f_{\rm He}) m_e}.
\end{equation}
The electron fraction and the helium fraction are defined by $x_e=n_e/n_H$ and $f_{\rm He}= \frac{y_p}{N_{\rm tot}(1-y_p)}$ respectively. Here $N_{\rm tot}=3.971$ and $y_p=0.21$, $\rho_{\rm CMB}= a_r T_{\rm CMB}^4$ (the radiation constant $a_r=7.565\times 10^{-16}$ J-m$^{-3}$-K$^{-4}$) and the Thomson scattering cross-section $\sigma_T=6.652\times 10^{-29}$ m$^2$. Coefficients $\alpha_{B}$, $\beta_{B}$ and $\gamma_e$ in equation (\ref{eq:ioni1}) are defined as \cite{Peebles:1968ja}
\begin{eqnarray}
	\alpha_B & = & F\times 10^{-19}\times \frac{4.309 \times (T_b/10^4)^{-0.616}}{1+0.670\times(T_b/10^4)^{0.530}} \,\, [{\rm m}^3 \, {\rm s}^{-1}], \\
	\beta_B & = & \alpha_B \left(\frac{2\pi m_e k_B T_{\rm CMB}}{h_{\rm pl}^2}\right)^{3/2}\exp\left(\frac{E_{21}}{k_B T_{\rm CMB}}\right)\, [ {\rm s}^{-1}]\, ,\\
	\gamma_e & = & 0.291\times10^{-7} \times \left(\left|\frac{E_{1s}}{k_B T_b}\right|\right)^{0.39} \, \frac{\exp(-|E_{1s}/ k_B T_b|)}{0.232+|E_{1s}/ k_B T_b|} 
	\nonumber \\	& & \times [{\rm cm}^3 {\rm s}^{-1}].
\end{eqnarray}
Here $\alpha_B$ is the Case-B recombination coefficient, $\beta_B$ is the photoionization rate and $\gamma_e$ is the collisional coefficient. Other parameters used are: $E_{21}=E_{1s}-E_{2s}=-10.2$ eV is energy of Ly$\alpha$ wavelength photon, $E_{1s}=-13.6$ eV, and $E_{2s}=-3.4$ eV. Coefficient $D$ represents the probability of a hydrogen atom in the state $n=2$ decaying to the ground state before photoionization can occur. It is defined as \cite{Peebles:1968ja}
\begin{eqnarray}\label{eq:Dfactor}
	D=\frac{1+K_H\Lambda_{2s,1s} n_{\small B} (1-x_e)}{1+K_H \, n_{\small B} (1-x_e)+K_H n_B \beta_B (1-x_e)}
\end{eqnarray}
where $K_H= \lambda_{H_{2p}^3}/8\pi H$, $\lambda_{H_{2p}}=121.568$ nm and  $\Lambda_{2s,1s}= 8.22~ {\rm s}^{-1}$ is the hydrogen two photon decay rate.
\subsection{Cooling effects}
We have also included cooling effects from the bremsstrahlung, collisional excitation, recombination and collisional ionization cooling. With these cooling effects, the evolution equation for the baryon temperature has following form
\begin{eqnarray}
	\frac{dT_b^{\rm cool}}{dz}=\frac{\Gamma_{\rm cool}}{1.5k_B\, n_{\small B}\,(1+z)H}, \label{eq:baryon_temp-3} 
\end{eqnarray}
where 
\begin{eqnarray}
	\Gamma_{\rm cool}=\frac{x_e n_H }{1.5 k_B}\left[\Theta x_e +\Psi (1-x_e)+\eta x_e+\zeta (1-x_e)\right] \label{eq:cooling-terms},
\end{eqnarray}
Unit of $\Gamma_{\rm cool}$ is $[\text{J}\, \text{m}^3\,\text{s}^{-1}]$.  Here $\Theta$, $\Psi$, $\eta$ and $\zeta$ are defined as (\cite{Fukugita:1994mm})
\begin{eqnarray}
	\centering
	\Theta& = & 5.84 \times 10^{-37}\, \left(\frac{T_b}{10^5}\right)^{0.5}\, , \nonumber \\
	\Psi & = & 7.50\times 10^{-31}  \left(1+ \left(\frac{T_b}{10^5}\right)\right)^{-1} \exp\left(-\frac{1.18}{T_b/10^5}\right), \nonumber \\
	\eta & = & 2.06 \times 10^{-36} \left(\frac{T_b}{10^5}\right)^{0.5} \left(\frac{T_b}{10^5}\right)^{-0.2} \left(1+ \left(\frac{T_b}{10^5}\right)^{0.7}\right)^{-1}, \nonumber \\
	\zeta & = & 4.02 \times 10^{-32}\left(\frac{T_b}{10^5}\right)^{0.5} \left(1+ \left(\frac{T_b}{10^5}\right)^{-1}\right)\exp\left(-\frac{1.58}{T_b/10^{5}}\right)\, . \nonumber \\
\end{eqnarray}
It is important to note here that all these functions are dependent on baryon temperature; hence they have a redshift dependence. The term on the right-hand side of the equation \eqref{eq:cooling-terms} has a positive sign; hence redshift lowers the baryon temperature.
\subsection{Magnetic heating}
\label{sec-2-sub-2.2}
In the present work, we consider a statistically homogeneous and isotropic random Gaussian, non-helical magnetic field background, which means we neglect any back reaction of the local matter evolution. During this period, we also assumes that universe is infinitely conducting. In this period $\vec{{\bf B}}(t, \vec{x})= \tilde{{\bf B}}(\vec{x})/a^2(t)$ on length scale larger than the magnetic Jeans length scale $\lambda_J^{-1}=k_J=\sqrt{8\pi \rho_m G}/{\rm v}_A$, where ${\rm v}_A=B_0/\sqrt{4\pi \rho_b}$ ($B_0$ is the comoving strength of magnetic fields and $\rho_b$ is the baryon density) \cite{Sethi:2004pe}. For the assumptions made above, power spectrum $\mathcal{P}_B(k)$ of the magnetic field is given by the relations \cite{Landau:1987, Brandenburg:2018ptt}
\begin{eqnarray}
	\langle \tilde{{\bf B}}^*_i({\bf k}) {\bf B}_j({\bf k}') \rangle =\frac{(2\pi)^3}{2}\, \delta_D ({\bf k}-{\bf k}')(\delta_{ij}-\hat{k}_i\hat{k}_j)\mathcal{P}_B(k)
\end{eqnarray}
where ${\bf B}^*_i({\bf k})$ is the Fourier component of ${\bf B}_i({\bf k})$ with a mode ${\bf k}$. Here power spectrum $\mathcal{P}_B(k)$ is defined as
\begin{equation}
	\label{eq:power-spec}
	\mathcal{P}_B(k)= 
	\begin{cases}
		\frac{(2\pi)^{n_B+5}}{\Gamma \left(\frac{n_B+3}{2}\right)} B_\lambda^2 \frac{k^{n_B}}{k_\lambda^{n_B+3}}  &,~~~ \text{for} ~k<k_D \\
		 0                          &,~~~  \text{for} ~k>k_D \\
	\end{cases}
\end{equation}
In the above equation, $B_\lambda$ is the strength of magnetic field smoothed at a length scale of $\lambda =k_\lambda^{-1}$ and $n_B$ is the magnetic field spectral index. Here $\lambda_D =k_D^{-1}$ is given by the following relation
\begin{equation} 
	\label{eq:mag-kd}
	\frac{k_D}{2 \pi{\rm Mpc}^{-1}} = \left[1.32\times 10^{-3} \left(\frac{B_0(t)}{1 {\rm nG}}\right)^2 \left(\frac{\Omega_b h^2}{0.02}\right)^{-1}\left(\frac{\Omega_m h^2}{0.15} \right)\right]^{\frac{-1}{(n_{\small B}+5)}}.
\end{equation}
Here $B_0(t)=B_0\, (1+z)^2$ is the comoving magnetic field strength. The spectral index $n_B=-2.9$ corresponds to the scale-invariant magnetic fields. In the present work, we have used values of $n_B>-2.9$ to avoid the infrared divergence of the magnetic fields \cite{Chluba:2015lpa}.

In the post recombination era, the radiative viscosity decreases quickly and at length scale $\lambda< \lambda_J$ (Jeans length scale), a decaying Magnetohydrodynamic (MHD) turbulence is generated due to the non-linear effects \cite{Jedamzik:1998kk, Subramanian:1997gi}. At these length scales, magnetic fields decay by two processes, (i) ambipolar diffusion and (ii) turbulent decay. The evolution equation of the magnetic energy density is given by 
\begin{eqnarray}
	\label{eq:magen-1}
	\frac{d}{dt}\left(\frac{|{\bf B}|^2}{8\pi}\right)=-4 H(t) \left(\frac{|{\bf B}|^2}{8\pi}\right) -\Gamma_{\rm ambi} -\Gamma_{\rm tur}.
\end{eqnarray}
We have followed the calculations done in \cite{Sethi:2003vp, Sethi:2004pe} for the ambipolar and turbulent processes. In our work, ambipolar decay is given by
\begin{eqnarray}
	\label{eq:ambi-1}
	\Gamma_{\rm ambi} = \frac{\rho_n}{16 \pi^2 \gamma \rho_b^2 \rho_i} |(\nabla \times {\bf B})\times {\bf B}|^2,
\end{eqnarray}
where $\rho_n$, $\rho_b$ and $\rho_i$ are the mass density of the neutral, ionized atoms and the baryon densities, respectively. In equation (\ref{eq:ambi-1}), $\gamma$ is given by (\cite{Schleicher:2008hc})
\begin{eqnarray}
	\gamma = \frac{\frac{1}{2}n_{\rm H} \langle \sigma v \rangle_{{\rm H}^+, {\rm  H}}+\frac{4}{5}n_{\rm He}\,\langle \sigma v \rangle_{{\rm H}^+, {\rm He}}}{m_{\rm H} \, (n_{\rm H}+4 n_{\rm He})}\,. 
\end{eqnarray}
Here $\langle \sigma v \rangle_{{\rm H}^+, {\rm  H}}$ and $\langle \sigma v \rangle_{{\rm H}^+, {\rm He}}$ is defined as (\cite{Schleicher:2008aa})
\begin{eqnarray}
	\langle \sigma v \rangle_{{\rm H}^+, {\rm  H}} & = & 0.649\, T_b^{0.375}\times 10^{-15}\, {\rm m}^3 \, s^{-1}, \nonumber \\
	\langle \sigma v \rangle_{{\rm H}^+, {\rm He}} & = &  [1.424+7.438\times 10^{-6} T_b-6.734\times 10^{-9} T_b^2]\nonumber \\
	&\times &  10^{-15}\, {\rm m}^3 \, s^{-1} .
\end{eqnarray}
Magnetic energy decay by MHD turbulence is given by \cite{Sethi:2004pe}
\begin{eqnarray} 
	\Gamma_{{\rm tur}} =  \frac{B_0^2(t)}{8\pi}\frac{3m}{2}\frac{\left[\ln\left(1+t_{\rm eddy}/t_i\right)\right]^m\, H(t)}{\left[\ln\left(1+t_{\rm eddy}/t_i\right)+\ln(t/t_i)\right]^{m+1}}\, , \label{eq:decay-1}
\end{eqnarray}
where $t_{\rm decay}=k_{D}^{-1}/{\rm v}_{\rm A}$ is the dynamical timescale and $t_i=1.248\times10^{13}$ s. $m$ is given by $m=2(n_{\small B}+3)/(n_{\small B}+5)$ (\cite{Olesen:1996ts, SHIROMIZU:1998ts, Jedamzik:1998kk, Christensson:2001ma, Sethi:2004pe}).
The decay of magnetic fields via ambipolar and turbulent process heats the baryons and  hence the contribution from the magnetic heating to the baryon temperature is given by
\begin{eqnarray}
	\frac{dT_{b}^{\rm mag}}{dz} = -\frac{2}{3 k_B n_H (1+z) H}(\Gamma_{\rm ambi} +\Gamma_{\rm tur})
\end{eqnarray}
The heating of the baryons as we move from high to low redshift is represented by the negative sign in the preceding equation.
\begin{figure*}
\hspace*{-0.3cm}
\subfloat[]{\includegraphics[width=0.40\linewidth, keepaspectratio]{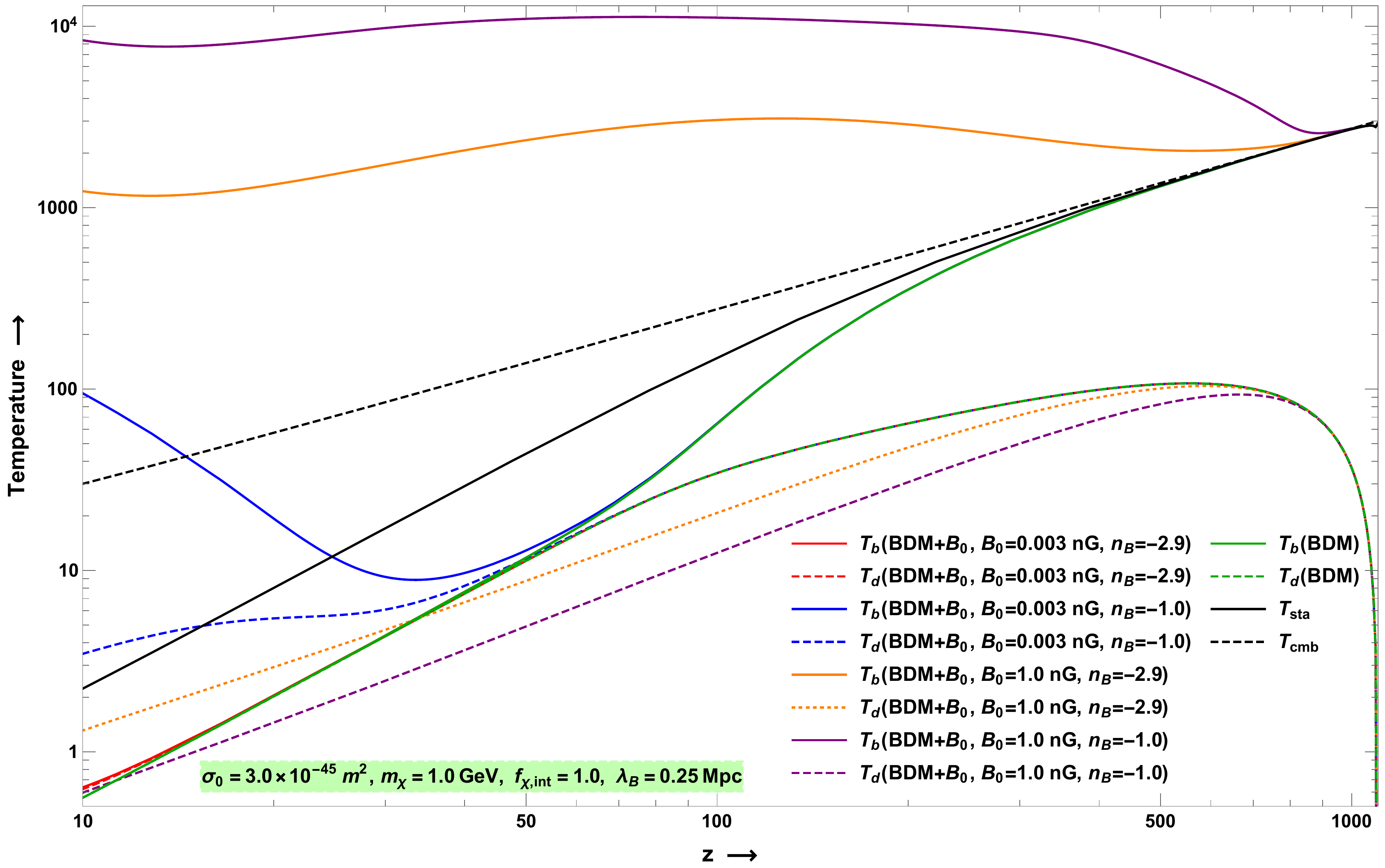}
	\label{tem-evolution-BDM-B0-only}}
\hspace*{0.5 cm}
\subfloat[]{\includegraphics[width=0.40\linewidth, keepaspectratio]{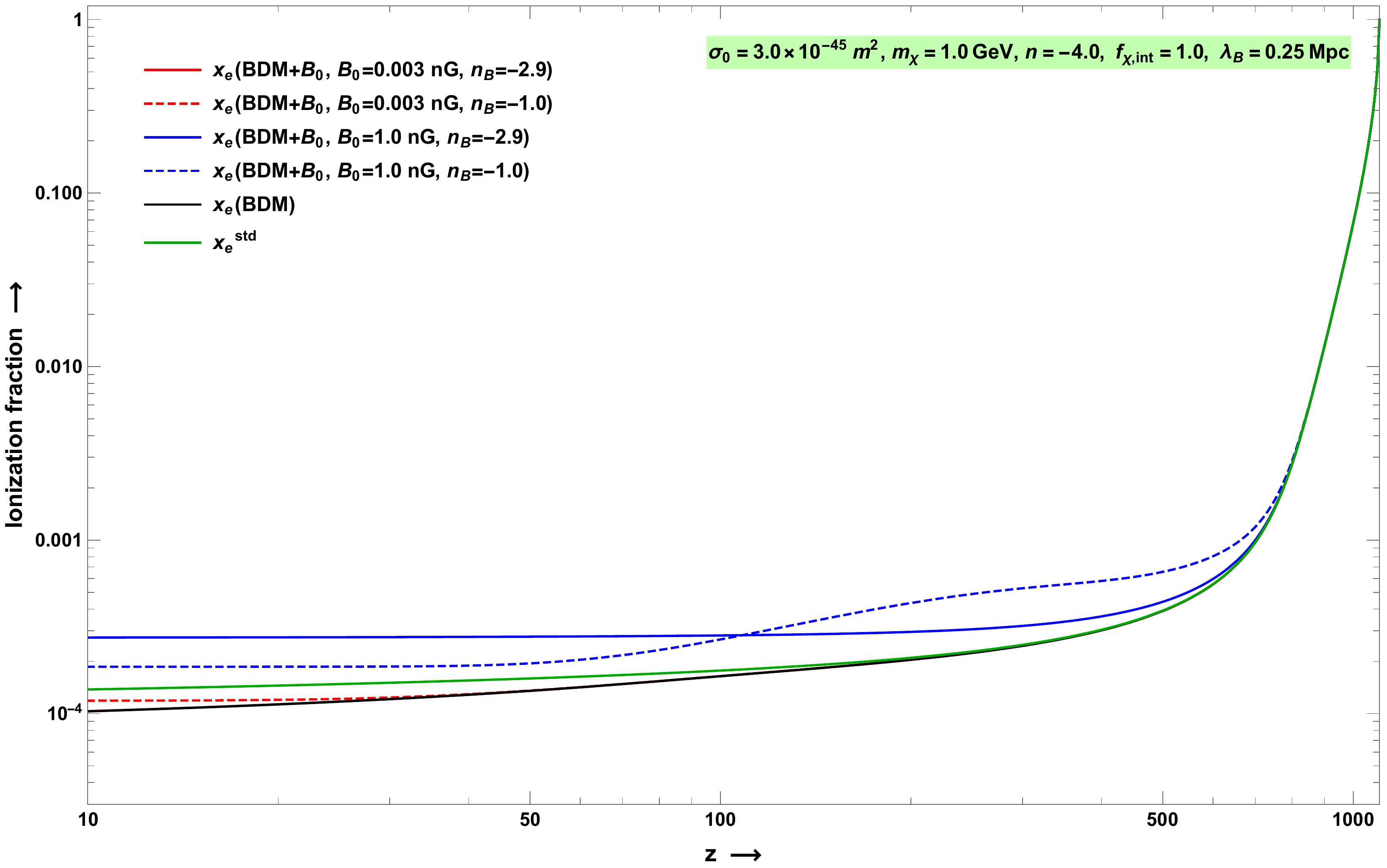}
\label{xe-evolutio-BDM-B0-only}} 
\caption{\textbf{BDM+PMF:} 
Evolution of baryon temperature and ionization fraction with respect to redshift, $z$. Magnetic fields contribute via ambipolar diffusion and magneohydrodynamic turbulent decay. The color description is given in the plots. We note here that we have considered one on one Dark matter and baryon interactions (means all Dark matter particles are interacting with the baryons), and $f_{\chi_{,\rm in}}=1$.}
\label{fig:BDM-B0-net}
\end{figure*}
\begin{figure*}
\hspace*{-0.3cm}
\subfloat[]{\includegraphics[width=0.40\linewidth, keepaspectratio]{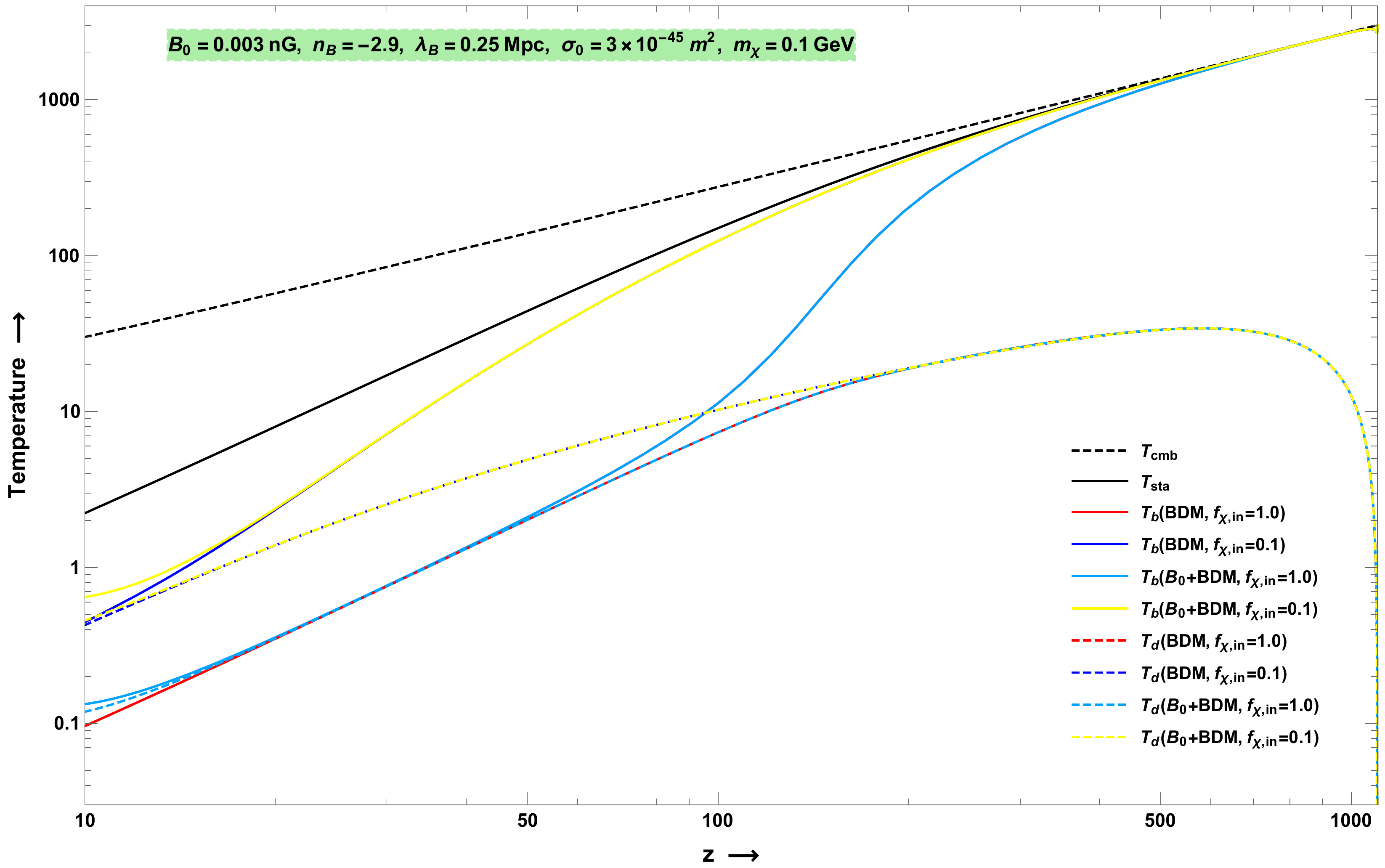}\label{BDM-B0-Tbfxint-z-Stan-B0003-BDM-sig-45-mdm0.1GeV}}
\hspace*{0.5 cm}
\subfloat[]{\includegraphics[width=0.40\linewidth, keepaspectratio]{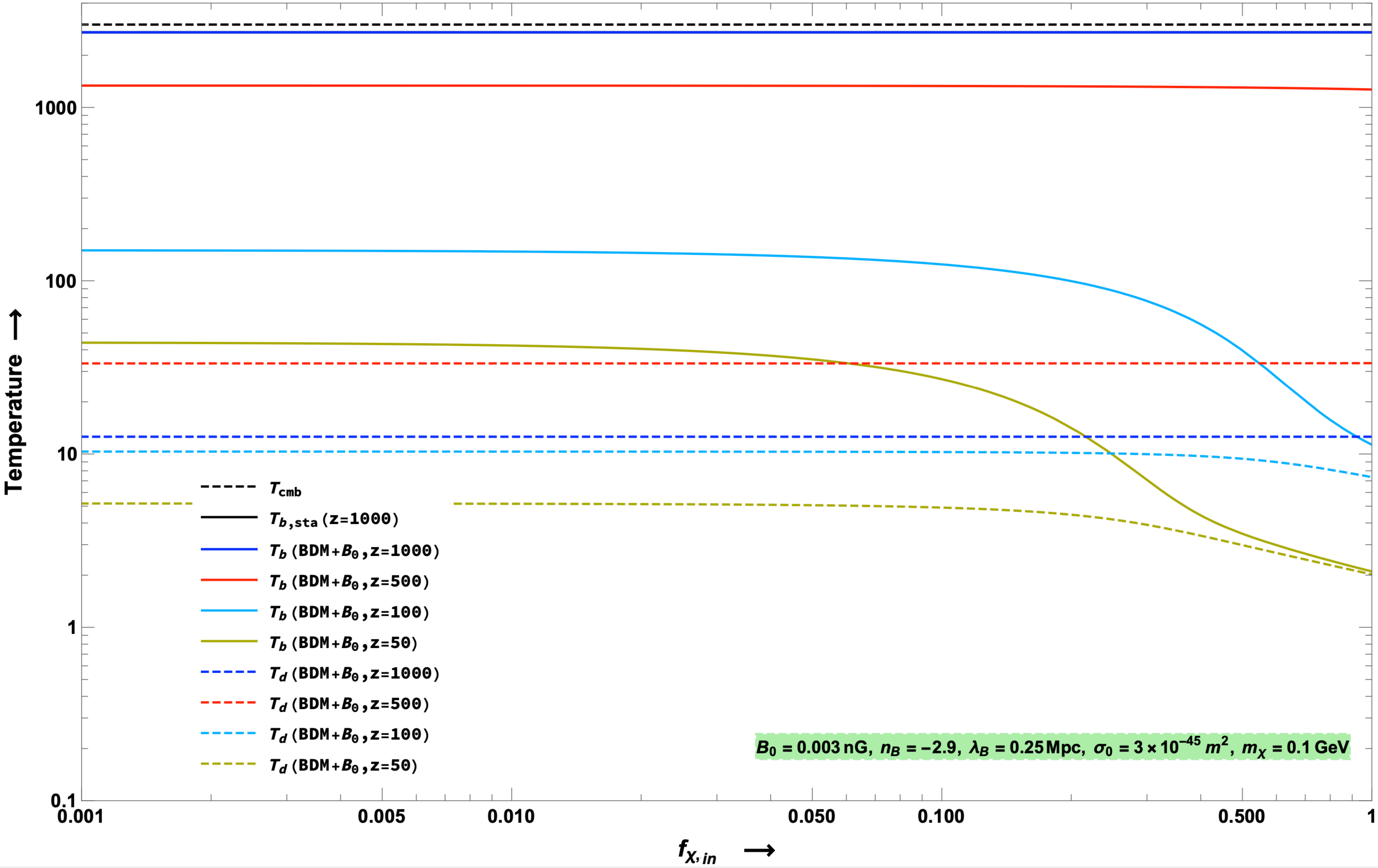}
\label{BDM-B0-Tbz-fxint-Stan-B003-BDM-sig-45-mdm0.1GeV}} 
\caption{\textbf{PMF+BDM:} Evolution of baryon temperature and ionization fraction and DM with respect to redshift $z$. Color description is given in the plots. In the left panel, we consider two values of $f_{\chi_{,\rm in}}=0.1$ and $1.0$ at given values of $B_0$, $\sigma_0$ and $m_\chi$.  On right panel, baryon temperature at a fix redshift, with respect to $f_{\chi_{,\rm in}}$. We use $\lambda_B=0.25$ Mpc.}
\label{fig:BDM+B0+fxint}
\end{figure*}
\subsection{Magnetic fields and the density evolution}
In the previous subsection, we have considered the effect of dissipation by ambipolar and turbulent decay of magnetic fields on the baryon temperature. It is known that these magnetic fields can generate density perturbations in the post-recombination epoch and later can collapse to form structures \cite{Wasserman:1978iw, Kim:1994zh, Subramanian:1997gi, Gopal:2003kr, Sethi:2004pe}. The density evolution equations of the baryons and DM, in presence of the magnetic fields at a length scale larger than the magnetic Jeans scale, are given by
\begin{eqnarray}
	&&\frac{\partial^2 \delta_b}{\partial t^2} + 2 H(t)\frac{\partial \delta_b}{\partial t}- 4\pi G (\rho_d \delta_d + \rho_b \delta_b) = S(t),	\label{eq:baryon-1} \\
	&&\frac{\partial^2 \delta_d}{\partial t^2} + 2 H(t) \frac{\partial \delta_d}{\partial t} - 4\pi G (\rho_d \delta_d + \rho_b \delta_b) = 0,	\label{eq:cdm-1}
\end{eqnarray}
where, $\rho_d$ is the cold dark matter density. $\delta_{\rm b}$ and $\delta_{\rm d}$ are the density contrast of the baryon and cold dark matter, respectively. In equation \eqref{eq:baryon-1}, the Lorentz force averaged over the coherence length scale of the magnetic fields are represented by the quantity $S(t)$ and it is defined as \cite{Minoda:2017iob}
\begin{eqnarray}
	S(t, {\bf x}) = \frac{\nabla \cdot \left[(\nabla \times \mathbf{B}(t,\mathbf{x})) \times \mathbf{B}(t,\mathbf{x})\right]}{4 \pi \rho_\mathrm{b} (t) a^2(t)}, \label{eq:source_b-1}
\end{eqnarray}
The absence of the source term in the DM equations from the magnetic fields signifies the absence of direct coupling of DM with the magnetic fields. In the range of interest of redshift $10< z< 1100$, density perturbation $\delta_b \leq 10^{-6}-10^{-5}$ is very small. In this scenario, analytic solution of above equations for $\delta_b$ gives (\cite{Minoda:2017iob}:
\begin{eqnarray}
	\delta_b = \frac{2S(t)}{15H^{2}(t)}
	\left[ \left\{ 3A \right. \right.
	+ 2A^{-\frac{3}{2}} - \left. 15 \ln A \right\} 	\frac{\Omega_b}{\Omega_m} + 15 \ln A\nonumber \\ 
	+ 30 \left(1 - \frac{\Omega_b}{\Omega_m}\right)
	A^{-\frac{1}{2}}
	- \left. \left(30-\frac{25 \Omega_b}{\Omega_m}\right)\right].
	\label{eq:delta_b-1}
\end{eqnarray}
The scale factor at the time of recombination is given by $a_{\rm rec}\propto (1+z_{\rm rec})^{-1}$ and quantity $A=a/a_{\rm rec}$. The additional contributes from the density fluctuation generated by the magnetic fields to the baryon temperature is given by $dT_b^\delta/dt =-T_bd(ln(1+\delta_b))/dt$ reduced to following equation
\begin{eqnarray}
	\frac{dT_b^\delta}{dz} = \frac{T_b}{1+\delta_b}\frac{d \delta_b}{dz}. 
	 \label{eq:baryon_temp-del}
\end{eqnarray}
A positive sign in front of the equation indicates that gas cools due to the growing density fluctuations induced by the magnetic fields with increase of the redshift.
\begin{figure*}
	\hspace{-0.3 cm}
	\subfloat[]{\includegraphics[width=0.40\linewidth, keepaspectratio]{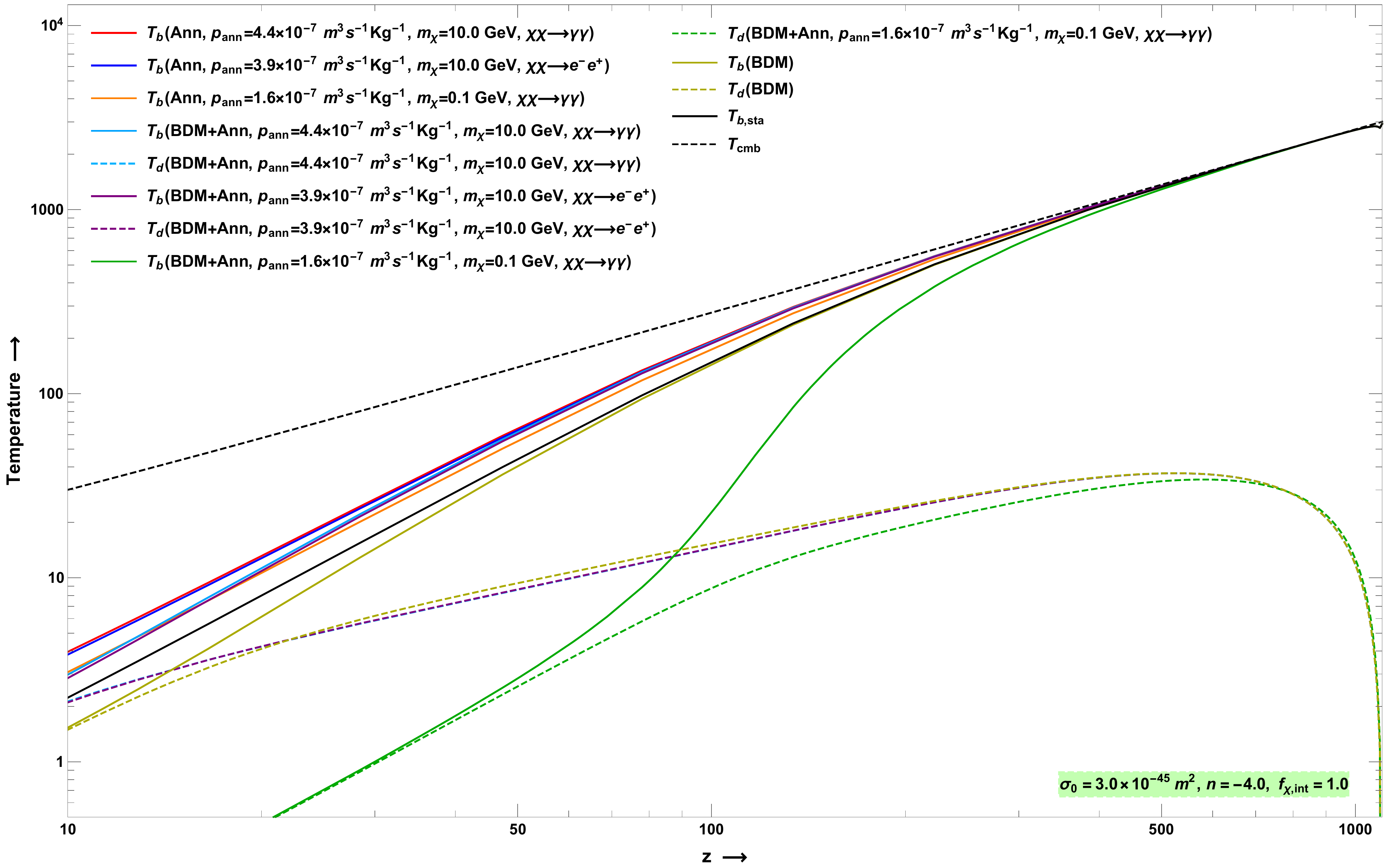}\label{fig:baryon-T-B003-1}}
	\hspace*{0.5 cm}
	\subfloat[]{\includegraphics[width=0.40\linewidth, keepaspectratio]{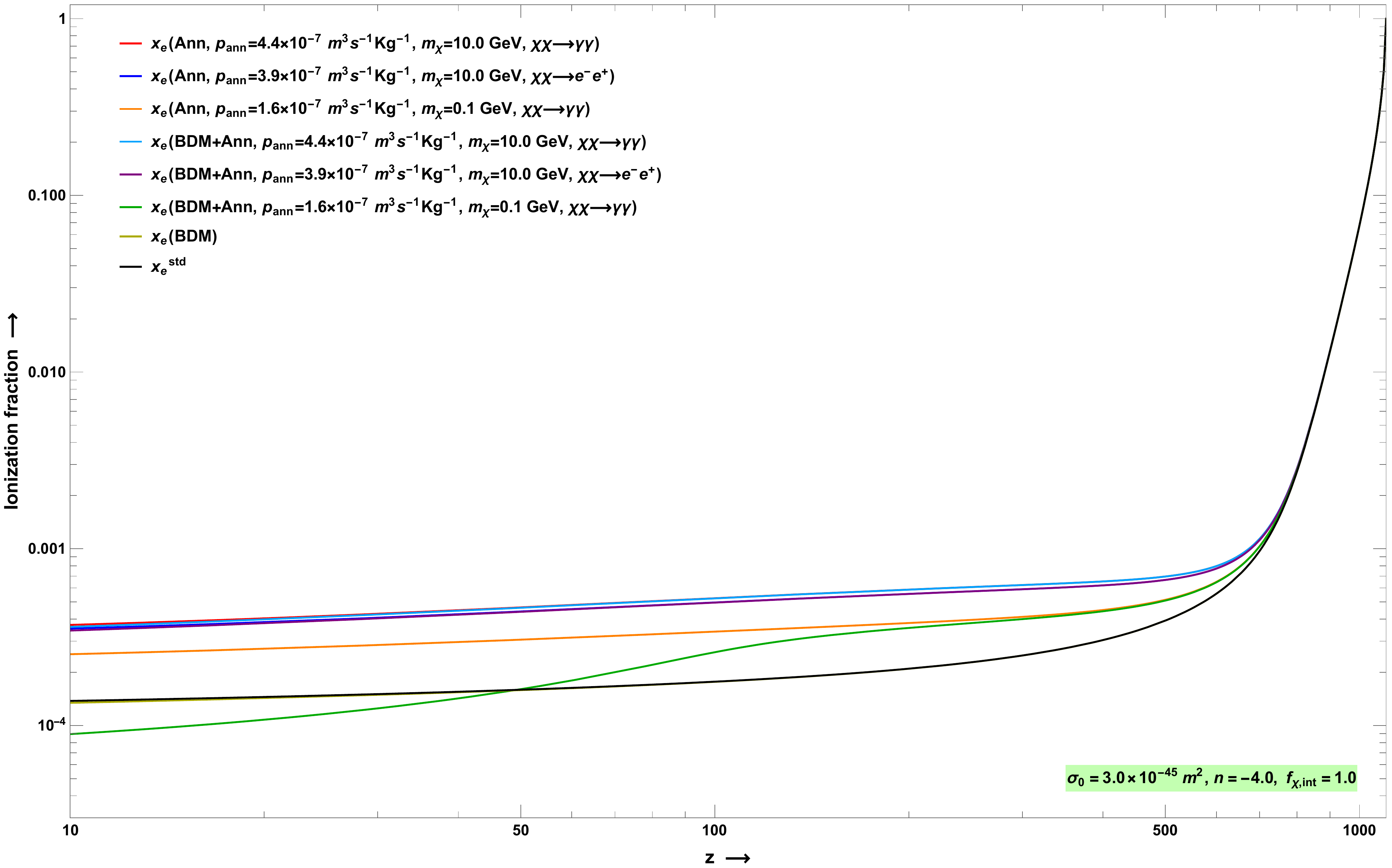}\label{fig:baryon-T-sig-42-1}} 
	\caption{\textbf{DM annhilation and BDM+Annihilation:} Evolution of $T_b$ and $x_{e}$ for the various combination of the parameters for the case of DM annhilation only and case when BDM+DM annihilation is considered. In the current figure, we set $\sigma_0=3.0\times 10^{-45}$ m$^2$, $n=-4.0$ for the scenario where we consider BDM interaction.}
	\label{fig:B-DM-Int-1}
\end{figure*}
\begin{figure*}
	\hspace{-0.3 cm}
	\subfloat[]{\includegraphics[width=0.40\linewidth, keepaspectratio]{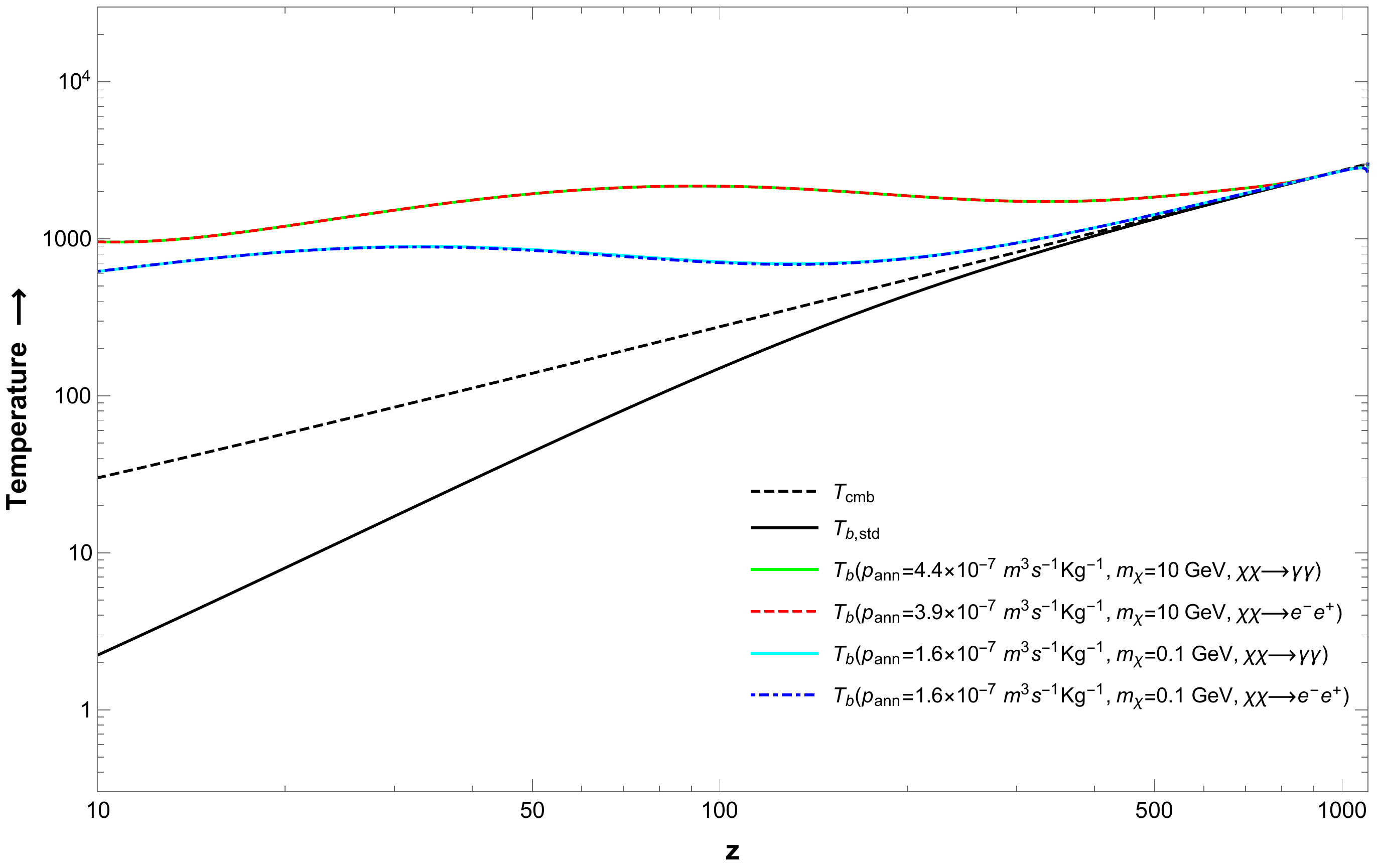}\label{fig:baryon-T-B003}}
	\hspace*{0.5 cm}
	\subfloat[]{\includegraphics[width=0.40\linewidth, keepaspectratio]{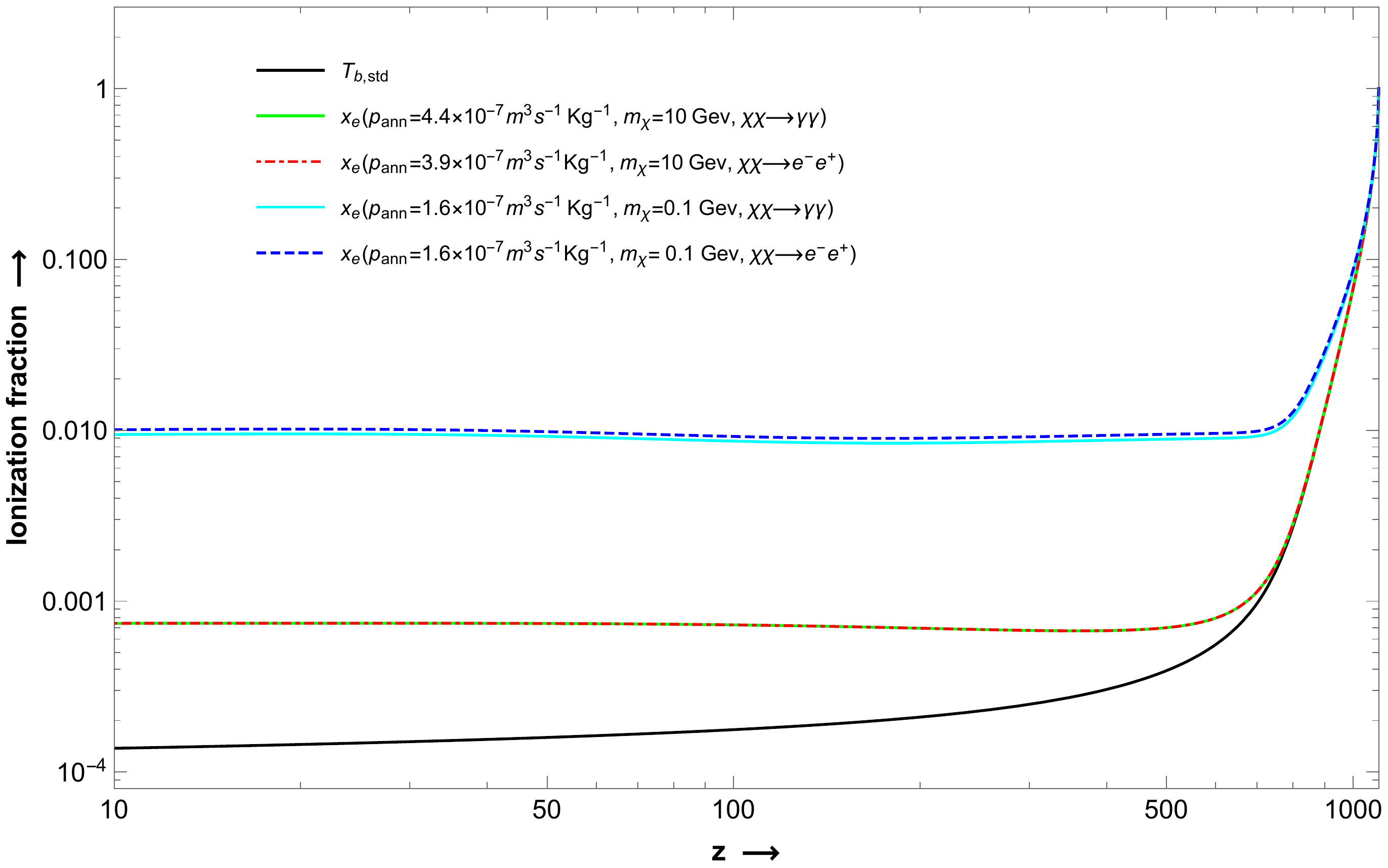}\label{fig:baryon-T-sig-42}} 
	\caption{\textbf{DM annhilation and PMF:}
		Evolution of $T_b$ at fixed values of $B_0=1 \rm{nG}$,$n_{B} = -1.0$ and $\lambda_B = 1.0$ Mpc and in presence of DM annhilation with many possible combinations of the channels and annhilation cross-sections is shown in figures (\ref{fig:baryon-T-B003}) and (\ref{fig:baryon-T-sig-42}), respectively.}
	\label{fig:B-DM-Int}
\end{figure*}
\begin{figure*}
	\hspace*{-0.3 cm}
	\subfloat[]{\includegraphics[width=0.40\linewidth, keepaspectratio]{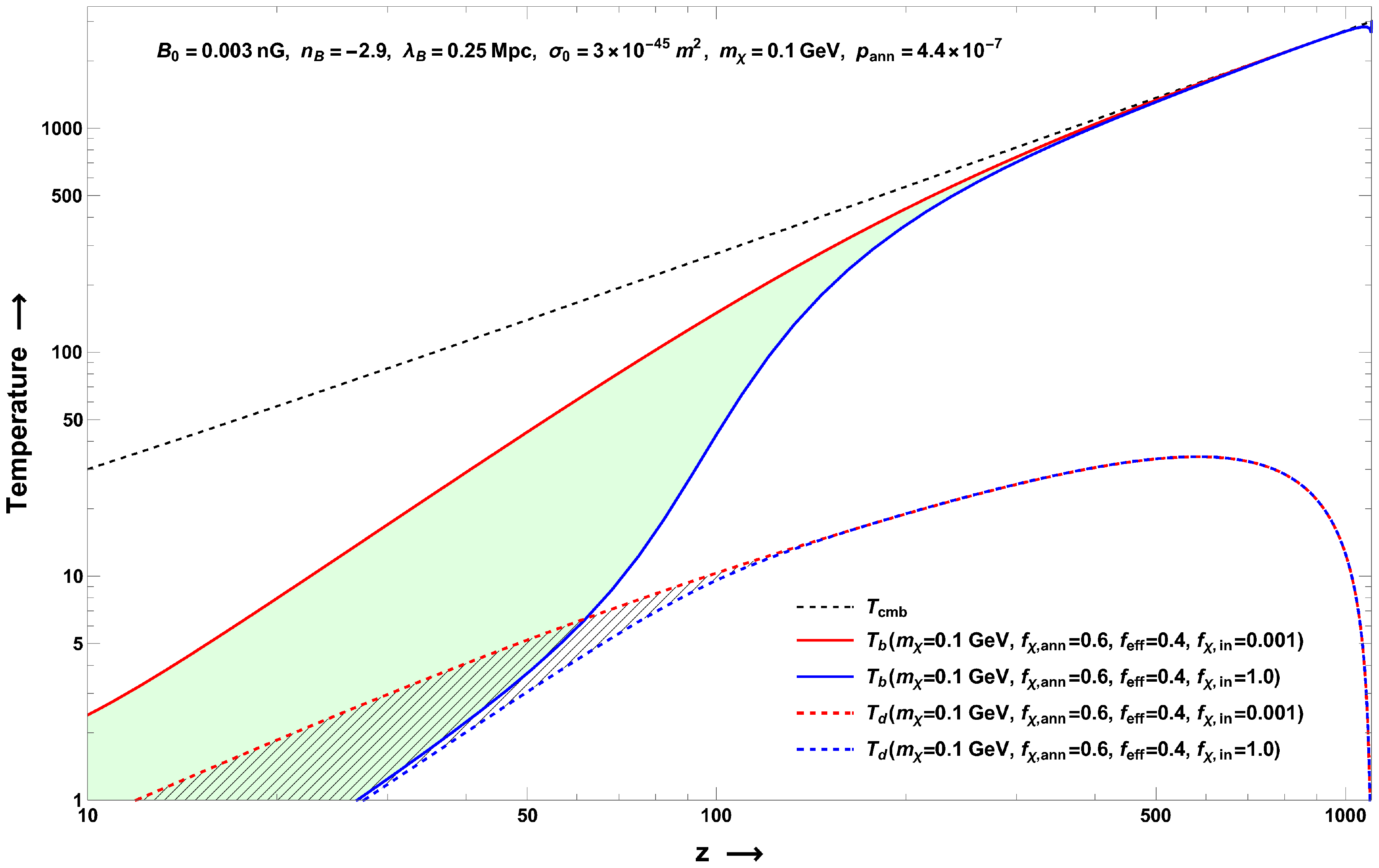}\label{fig:Tb-Td-max-min-fxint-plot-arun}}
	\hspace*{0.5 cm}
	\subfloat[]{\includegraphics[width=0.40\linewidth, keepaspectratio]{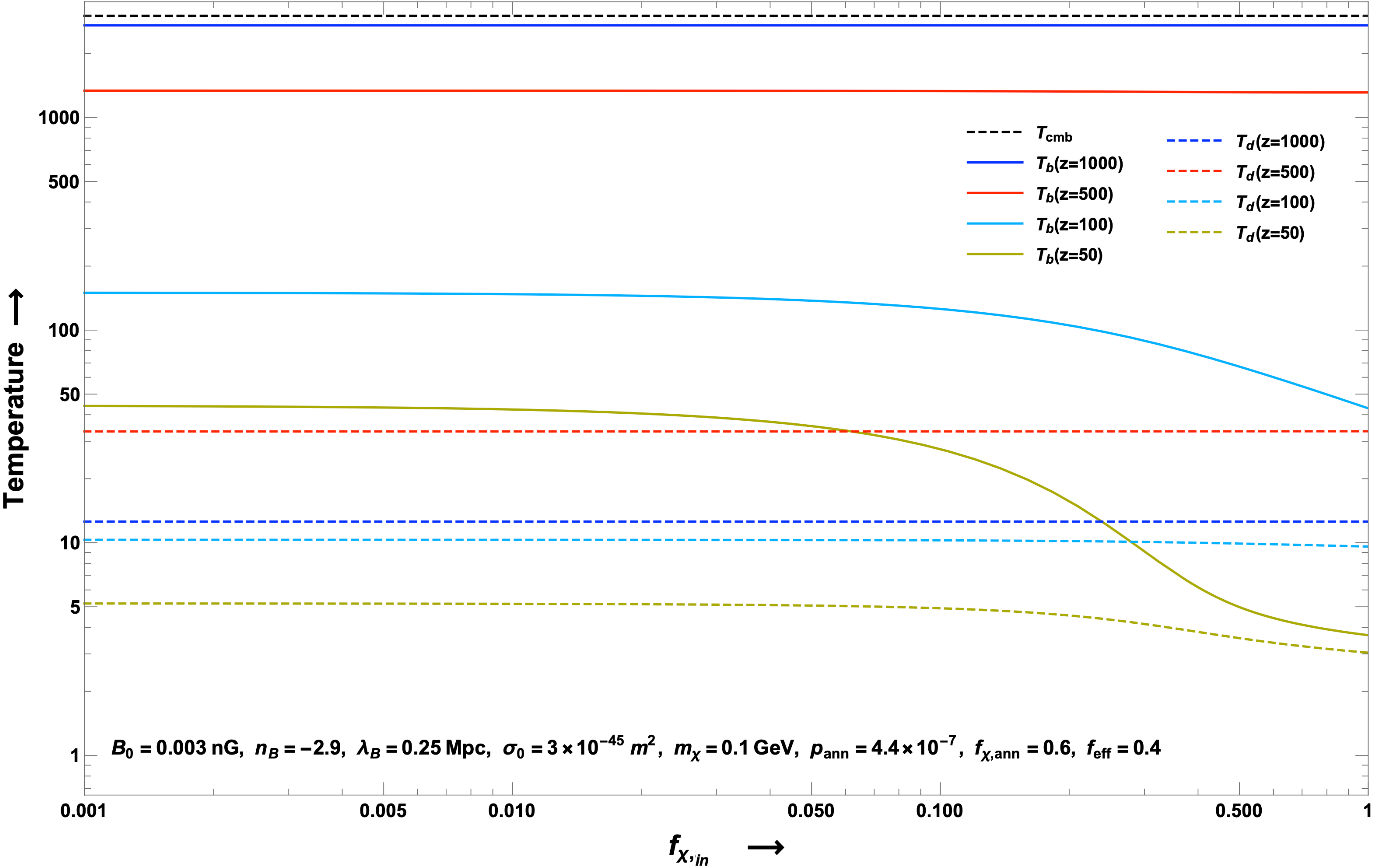}
	\label{fig:TB-Tdsig-vs-fxint}}
	\caption{\textbf{BDM+PMF+Annihilation}: Baryon and DM temperature for two values of $f_{\chi,{\rm int}}$ at fix values of $f_{\chi,{\rm ann}}$ and $f_{\rm eff}$. The light green color between the two T$_b$ curves and the hatched area between the two T$_d$ curves shows the area for for $f_{\chi,{\rm int}}=0.001$ and $1$. The results are consistent to the result obtained in \citep{Liu:2018uzy}.}
	\label{fig:detail-fxint-sig}
\end{figure*}
\begin{figure*}
	\hspace*{-0.4 cm}
	\subfloat[]{\includegraphics[width=0.40\textwidth, keepaspectratio]{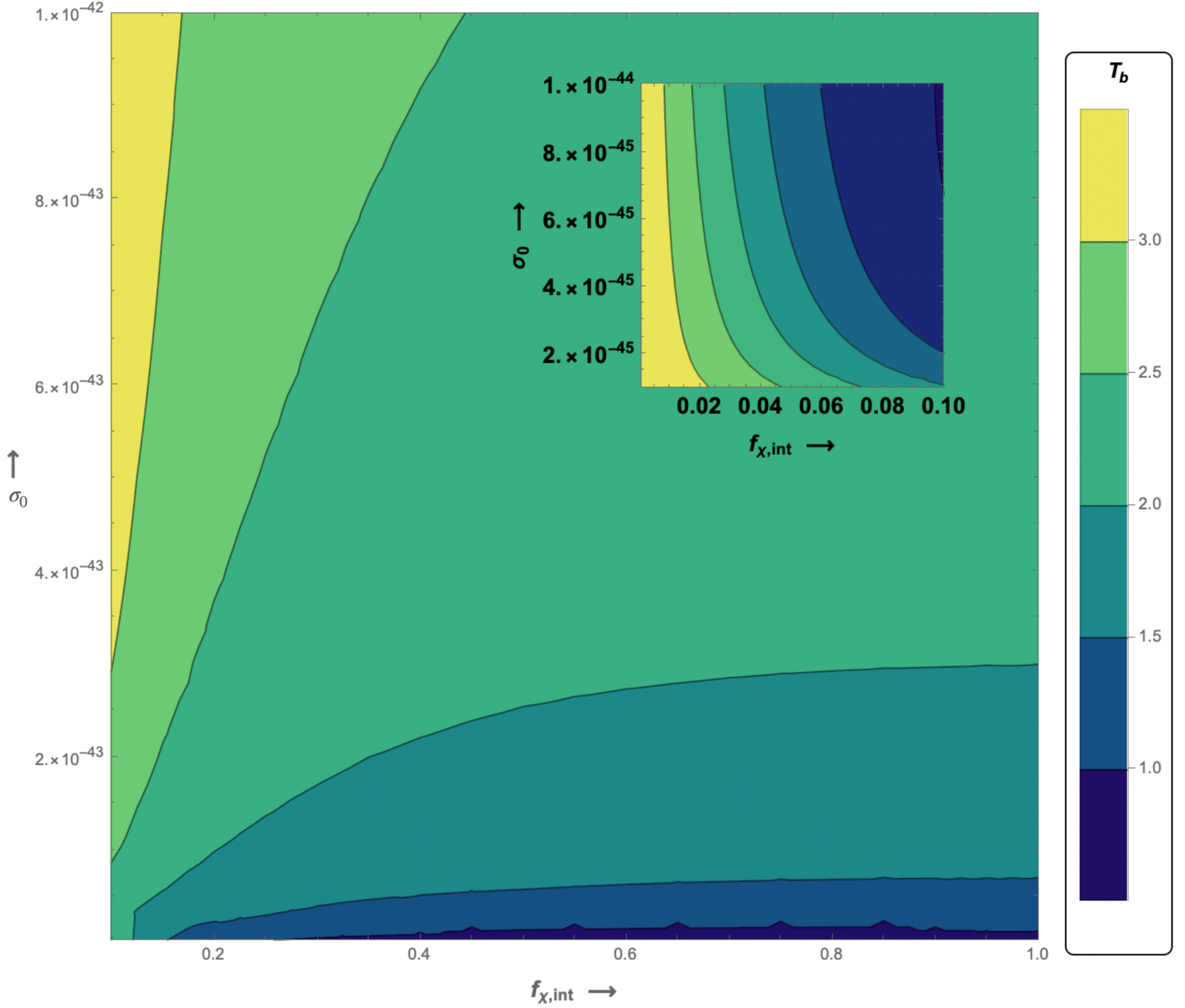}\label{fig:Tb-Td-vs-fxint}}
	\hspace*{0.5 cm}
	\subfloat[]{\includegraphics[width=0.40\textwidth, keepaspectratio]{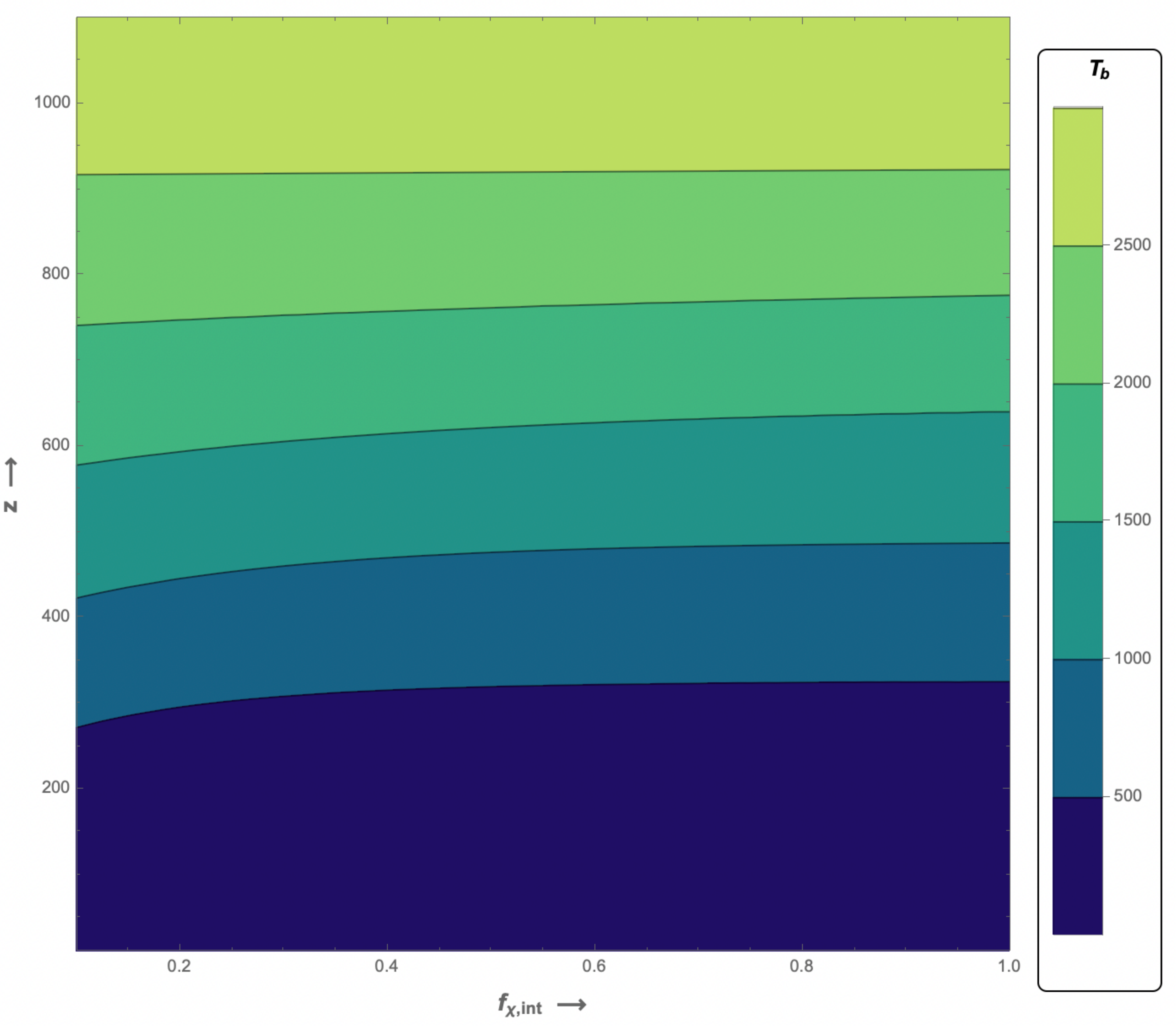}\label{fig:Tb-vs-fxint-contour}}
	\caption{\textbf{BDM+PMF+DM annihilation:} Here we have shown dependence of baryon temperature on the $f_{\chi,{\rm int}}$ and $\sigma_0$. In left figure, countour plot of $T_b$ in $\sigma_0-f_{\chi,{\rm int}}$ plain at redshift $z=20$ is given. In the right panel, we show the contour plot of $T_b$ in $z-f_{\chi,{\rm int}}$ plain for $\sigma_0=3.0\times10^{-45}$ m$^2$. In both cases, we set $B_0=0.003$ nG, $n_B=-2.9$, $\lambda_B=0.25$ Mpc, $m_\chi=0.1$ GeV, $p_{\rm ann}=4.4\times 10^{-7}$ m$^3$ sec$^{-1}$ Kg$^{-1}$, $f_{\chi, {\rm ann}}=0.6$ and $f_{\rm eff}=0.4$.}
 \label{fig:detail-fxint}
\end{figure*}
\begin{figure*}
	\hspace*{-0.3 cm}
	\subfloat[]{\includegraphics[width=0.40\linewidth, keepaspectratio]{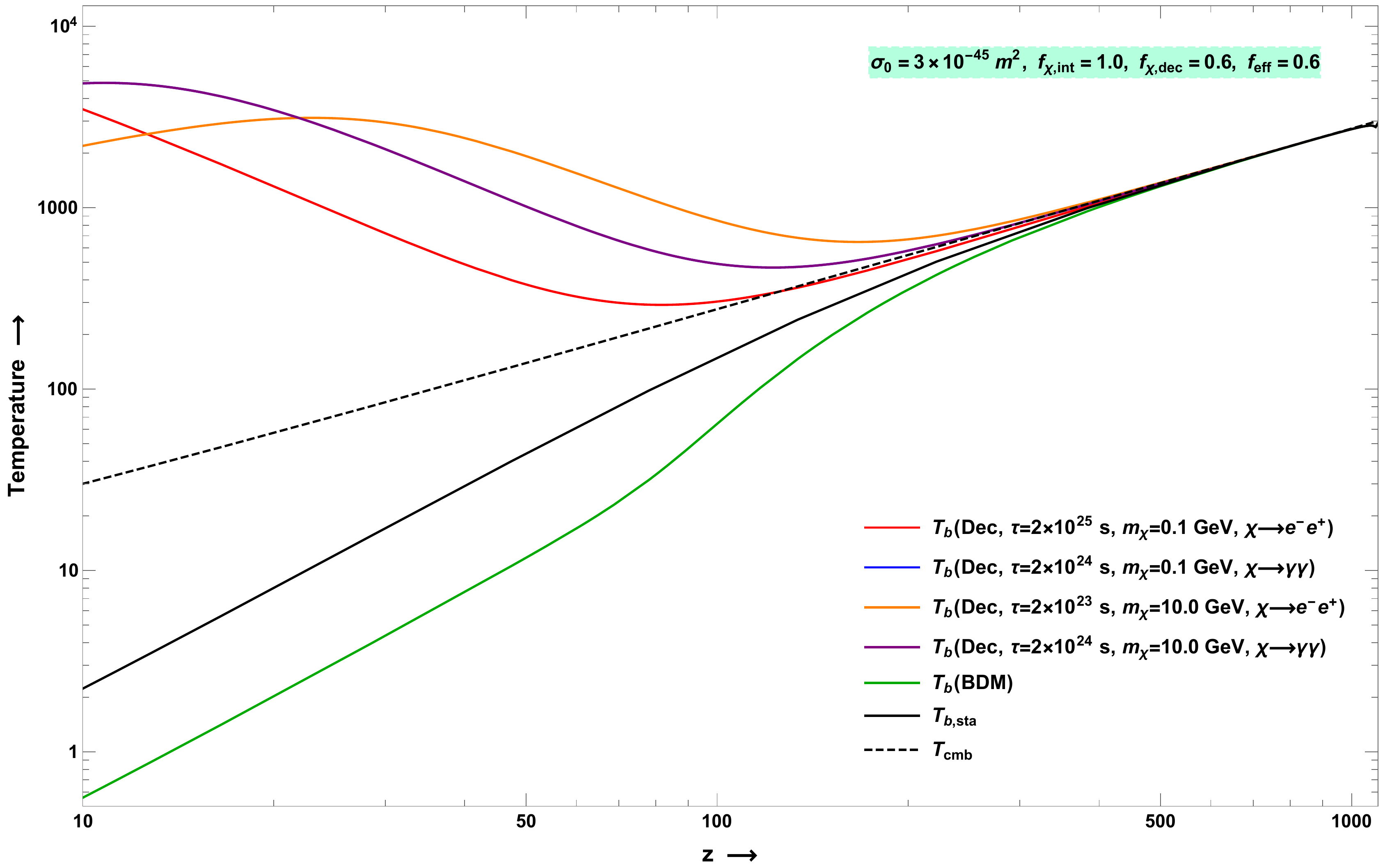}\label{fig:Td-B05-1}}
	\hspace*{0.5 cm}
	\subfloat[]{\includegraphics[width=0.40\linewidth, keepaspectratio]{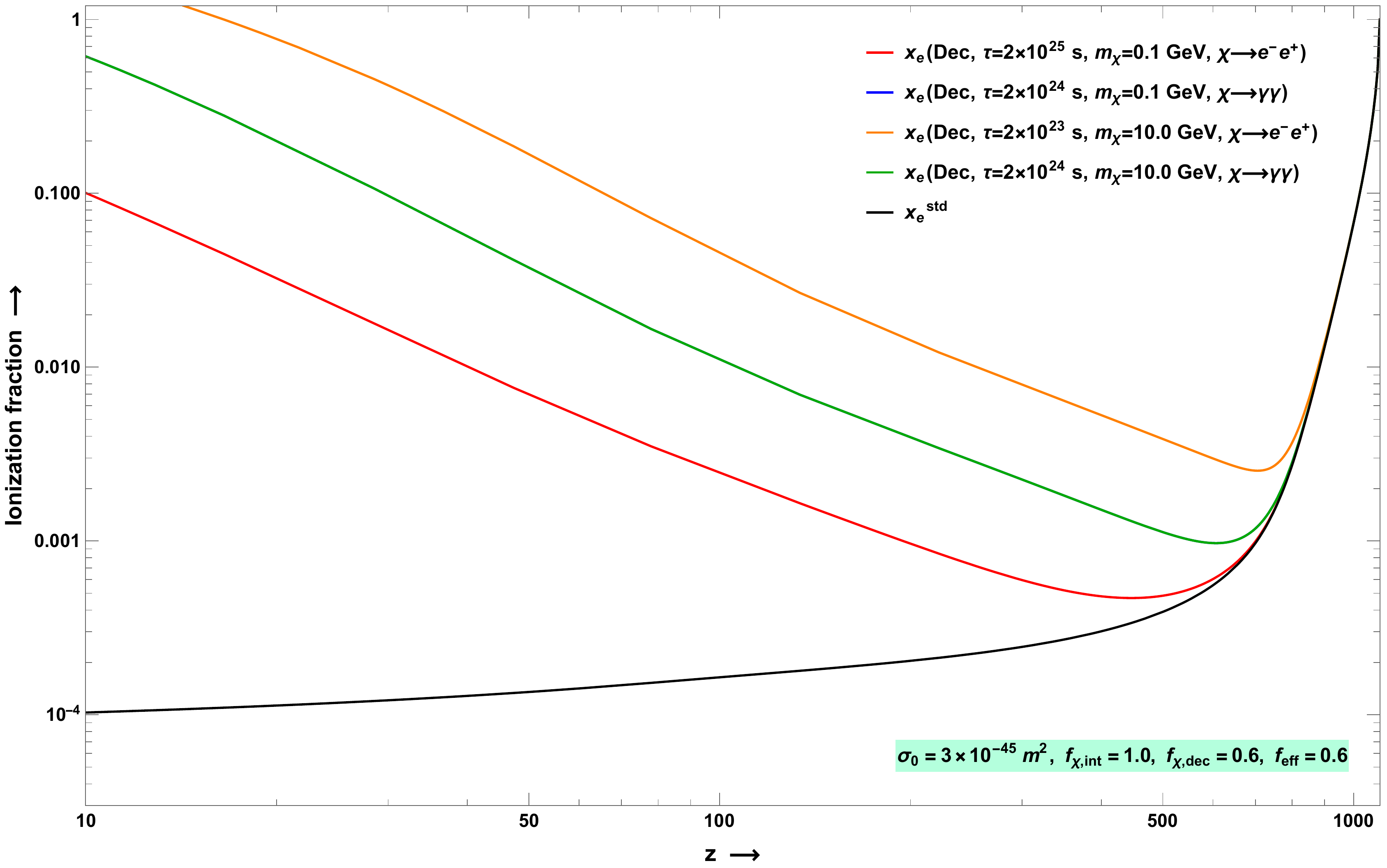}\label{fig:Td-sig042-1}}
	\caption{\textbf{DM decay only} Evolution of $T_b$ and $x_{e}$ for the various combination of the parameters. Here we compare baryon temperature for two DM decay channels with the standard baryon temperature evolution.}
	\label{fig:detail-1}
\end{figure*}
\begin{figure*}
	\hspace*{-0.3 cm}
	\subfloat[]{\includegraphics[width=0.40\linewidth, keepaspectratio]{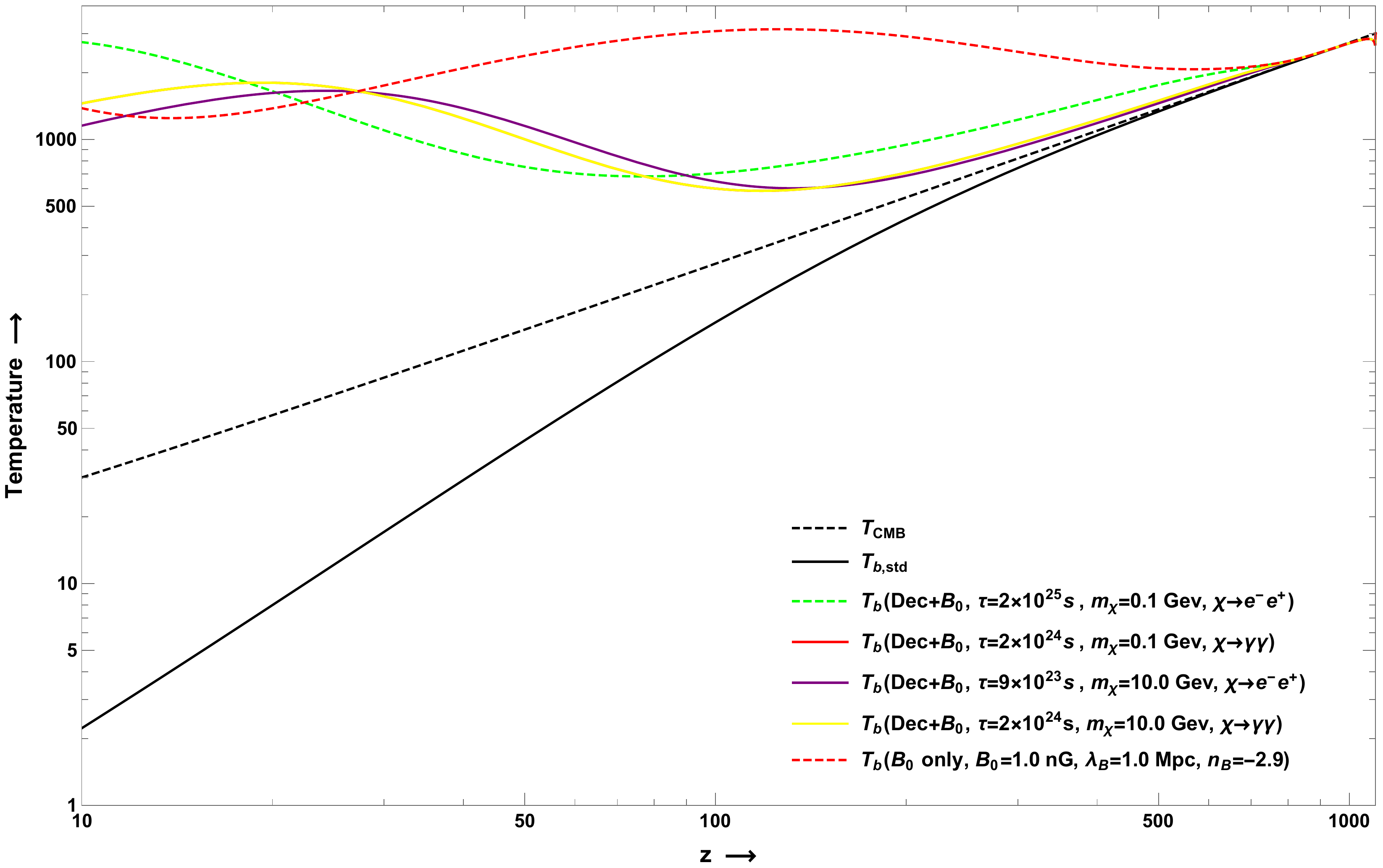}\label{fig:Td-B05}}
	\hspace*{0.5 cm}
	\subfloat[]{\includegraphics[width=0.40\linewidth, keepaspectratio]{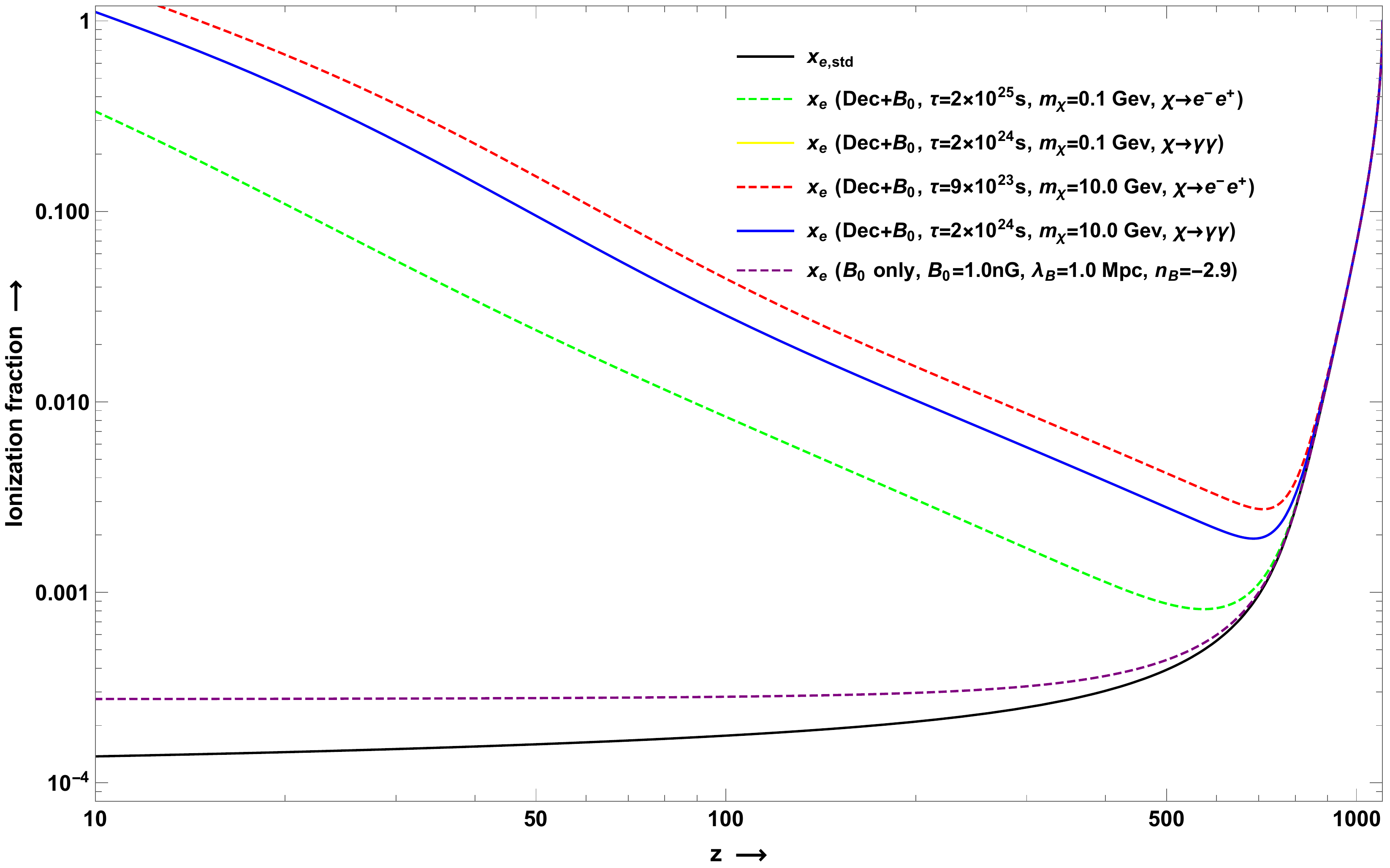}\label{fig:Td-sig042}}
	\caption{\textbf{DM decay+PMF:} Evolution of $T_b$ at fixed values of $B_0=1 \rm{nG}$, $n_{B} = -1.0$ and $\lambda_B = 1.0$ Mpc and in presence of DM decay with many possible combinations of the channels and decay time is shown in figures (\ref{fig:Td-B05}) and (\ref{fig:Td-sig042}) respectively.}
	\label{fig:detail}
\end{figure*}
\begin{figure*}
	\hspace*{-0.3 cm}
	\subfloat[]{\includegraphics[width=0.40\linewidth, keepaspectratio]{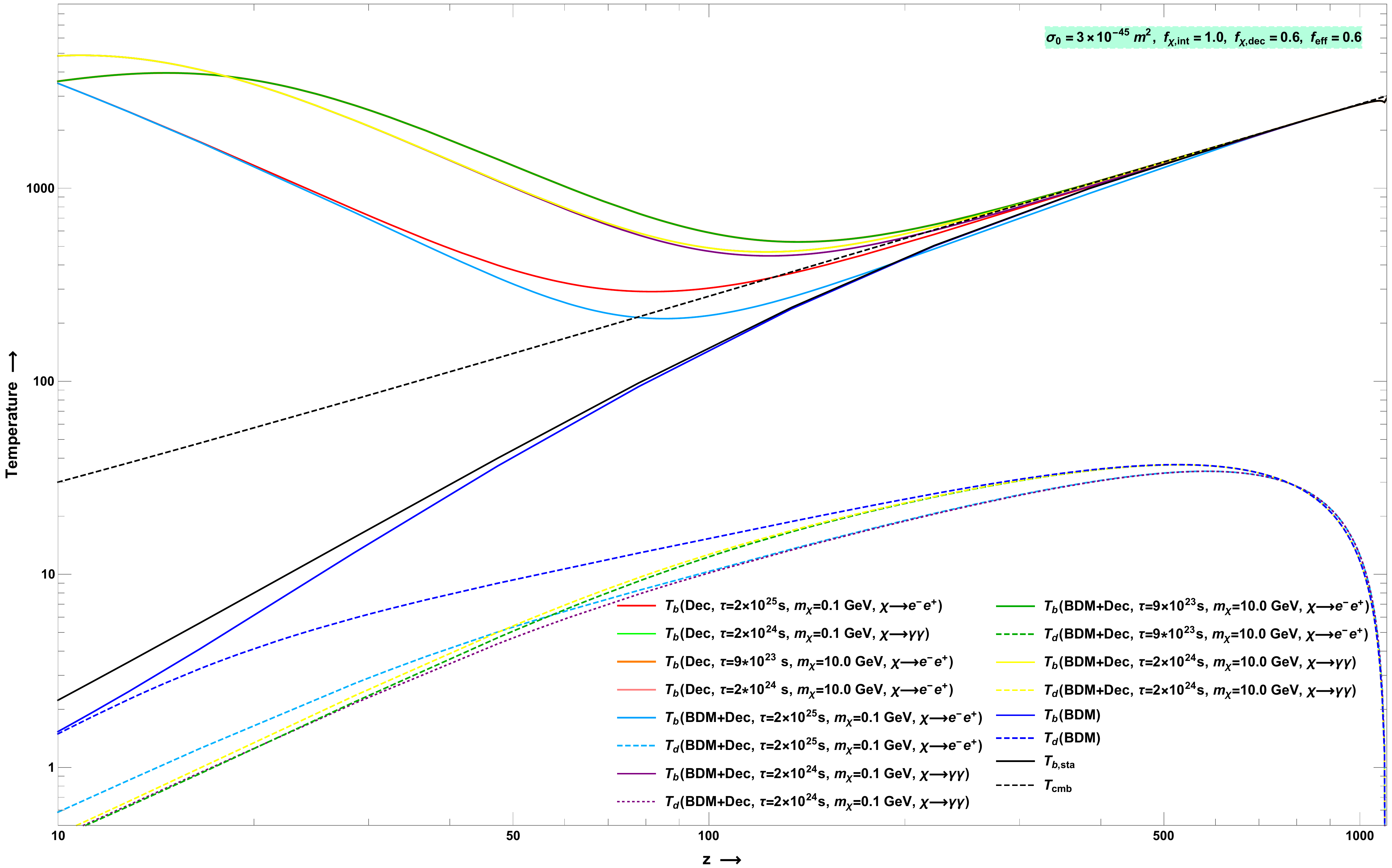}\label{fig:decBDM-tbtd}}
	\hspace*{0.5 cm}
	\subfloat[]{\includegraphics[width=0.40\linewidth, keepaspectratio]{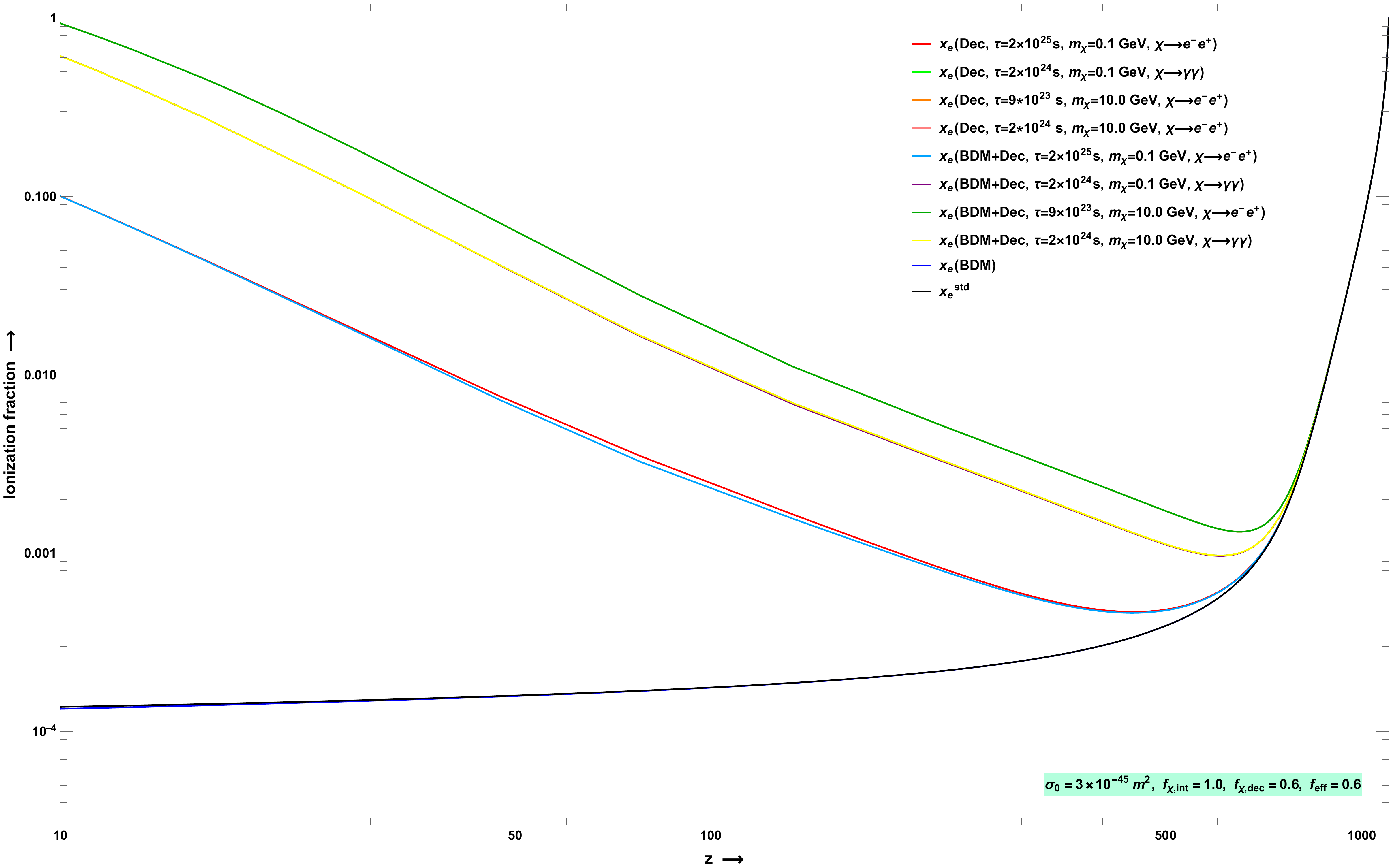}\label{fig:decBDM-xe-9}}
	\caption{\textbf{DM decay+BDM:} Evolution of $T_b$ at fixed values of $B_0=0.003 \rm{nG}$, $n_{B} = -1.0$ and $\lambda_B = 1.0$ Mpc and in presence of DM decat with many possible combinations of the channels and decay time is shown in figures (\ref{fig:Td-B05}) and (\ref{fig:Td-sig042}) respectively.}
	\label{fig:decBDM-9}
\end{figure*}
\begin{figure*}
	\hspace*{-0.3 cm}
	\subfloat[]{\includegraphics[width=0.40\linewidth, keepaspectratio]{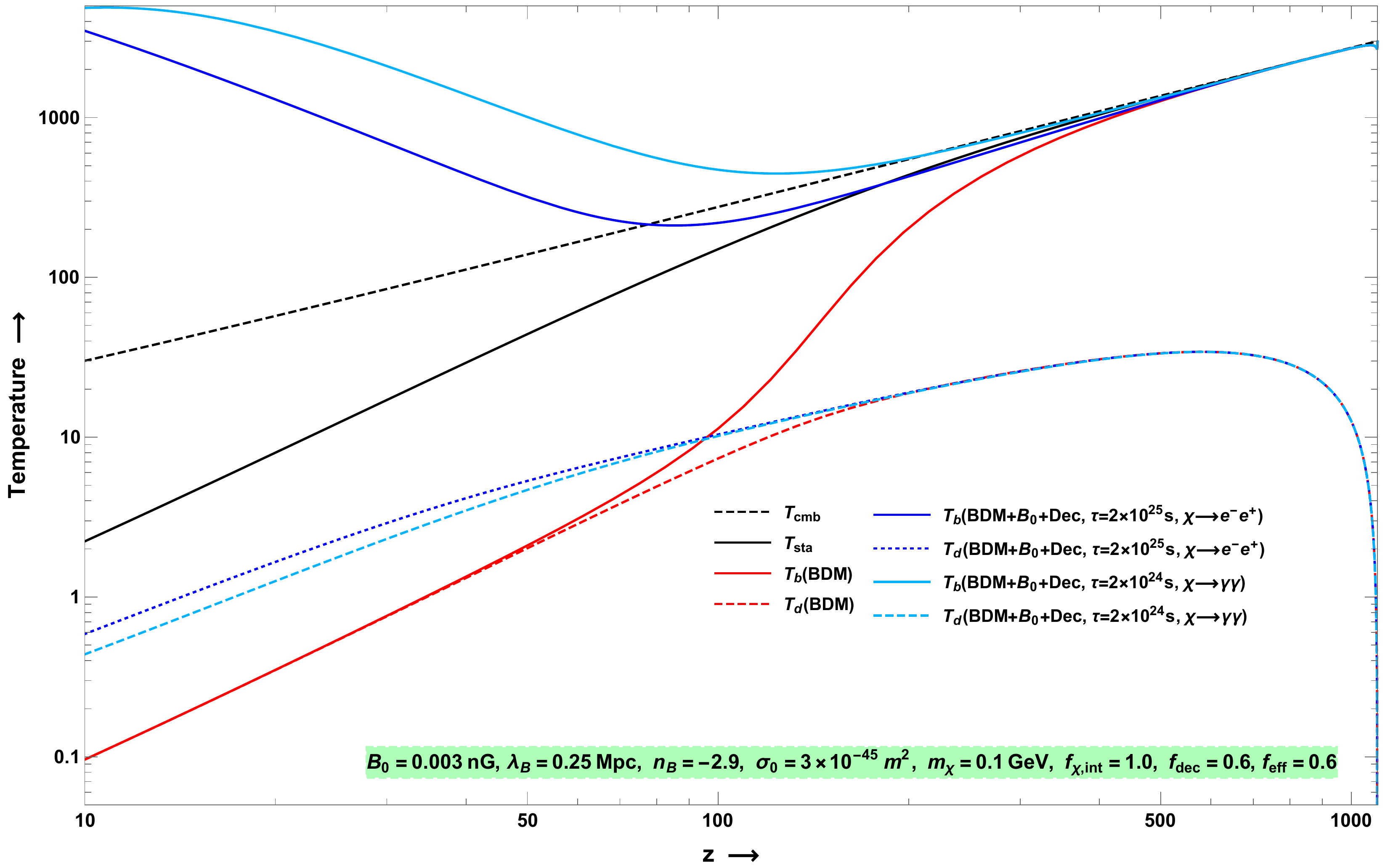}\label{fig:dec-tbtd}}
	\hspace*{0.5 cm}
	\subfloat[]{\includegraphics[width=0.40\linewidth, keepaspectratio]{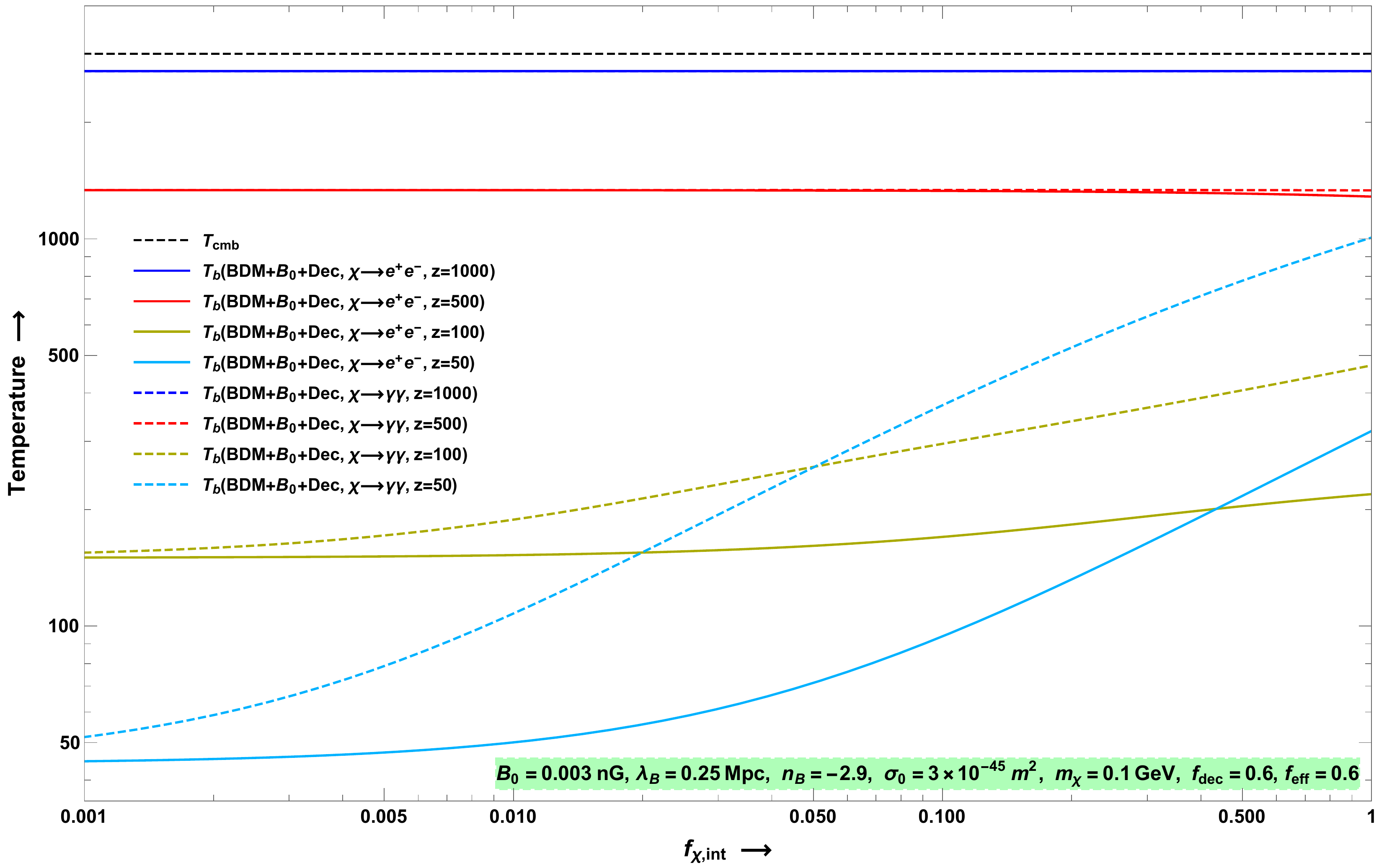}\label{fig:dec-tbtd-fxint}}
	\caption{\textbf{DM decay+PMF+BDM:} Evolution of $T_b$ at fixed values of $B_0=0.003 \rm{nG}$, $n_{B} = -1.0$ and $\lambda_B = 1.0$ Mpc and in presence of DM decat with many possible combinations of the channels and decay time is shown in figures (\ref{fig:Td-B05}) and (\ref{fig:Td-sig042}) respectively.}
	\label{fig:detail-2}
\end{figure*}
\subsection{Effect of the Baryon-DM interaction}
\label{sec-2-sub-2.3}
EDGES collaboration detected a global 21 cm absorption signal centered at 78 MHz at a redshift of $z\approx 17$, which is 2.5 times larger than the predicted from the standard $\Lambda$CDM model \cite{Bowman:2018yin}. After detecting this signal, various attempts have been made to explain the detected EDGES anomaly \cite{Moroi:2018vci, Fraser:2018acy}. Since DM is colder than the gas in the dark ages, gas particles can be cooled by elastic scattering of the baryons by the DM particles \cite{Fialkov:2014kta, Tashiro:2014tsa, Barkana:2018nd}. One of the crucial aspects of this consideration is to understand the behavior and structure of the DM particles,  whose nature is, however, still not known. The assumption that baryon and DM interact opens up a new area of research, where it is one of our best chances to understand these particles from a different angle. 

To explain the 21-cm absorption signal by EDGES collaboration at redshift $z\approx 17.2$, a non-standard Coulomb interaction cross-section $\sigma({\rm v}) =\sigma_0\, ({\rm v}/c)^{n}$ (here ${\rm v}$ is the relative velocity between the DM and baryons) between the DM and baryons is considered \cite{Bowman:2018yin, Barkana:2018nd}. The exponent $n$ in the above relation is a parameter and $n=-1$ correspondence to Yukawa potential, $n=-2$ is given for the case when the DM particles have an electric dipole moment and for a millicharge DM particles $n=-4$. Heat transfer between the DM and baryons occurs by two processes: i). one by the temperature difference between the baryons and the DM particles, ii). second one due to the drag force acting between the two particles. The second force act as resistive force and tend to become zero with time. The drag force per unit mass between the DM and baryons at the kinetic decoupling is quantified by the rate of change of the bulk velocity of the DM fluid with respect to the baryons. The net energy transfer due to the interaction from DM to baryon is therefore written as \cite{Dvorkin:2013cea, Tashiro:2014tsa, Munoz:2015bk, Munoz:2017qpy, Barkana:2018nd}
\begin{eqnarray}
	\label{eq:heat-tran-BDM}
&& \dot{Q}_b =  \frac{\rho_d\mu_b\, k_B\, c^{-n}\, \sigma_0 e^{-r^{2}/2}} {(m_d+m_b)^2}\left(\frac{k_B T_b}{m_b} + \frac{k_BT_d}{m_d} \right)^{\frac{n+1}{2}}\nonumber \,\\ 
&& \times \frac{2^\frac{5+n}{2} \Gamma\left(3+\frac{n}{2}\right)}{\sqrt{\pi}} (T_d -T_b) +\frac{\rho_d}{\rho_d + \rho_b}\frac{m_d\, m_b}{m_d+m_b}\, |{\rm v}\, \frac{d \textbf{v}}{dt}|, \nonumber \\
\end{eqnarray}
Similarly, we can write the expression for the energy exchange from baryons to DM is given by just exchanging $b\leftrightarrow d$. First-term in the above equation on the right-hand side represents the heat transfer due to the finite temperature difference between the baryons and the DM particles. However, the second term is given for the effects of drag force. In above equation (\ref{eq:heat-tran-BDM}) $\rho_m$ is the total matter mass density, and $m_b$, $m_d$ are the mass of the baryon and dark matter particles, respectively. The variables $r$ is defined as $r \equiv \text{v}/u_{\rm th}$ and $u_{\rm th}$ is given by: $u_{\rm th}^2= \left(\frac{k_b T_b}{m_b}+\frac{k_b T_d}{m_d}\right)$. Also $\mu_b\simeq m_H(n_H+4 n_{He})/(n_H+n_{He}+n_e)$. 
For large values of $n$, DM and baryons are coupled until the scattering rate is lower than the Hubble expansion rate of the universe. For example, when $n=0$, the decoupling time scale of the baryon-DM is given by $\tau_{\rm BDM}\equiv (a/R_\chi)(m_d+m_H)/m_d$. $R_\chi$ is the baryon-DM momentum exchange rate, and it is given in reference \cite{Xu:2018efh}. In this case, the temperature of the baryons and the DM particles remain the same when $H \tau_{\rm BDM}>1$. We have also assumed that the DM is non-relativistic and its velocity decreases only adiabatically with $v\propto (1+z)$ after the decoupling epoch. Therefore, the evolution equation of the baryons and the DM are following
\begin{eqnarray}
	\frac{dT_{b_{\rm BDM}}}{dz} & =  & - \frac{2\, \dot{Q}_b}{3 k_B (1+z) H} \\
	\frac{dT_{d_{\rm BDM}}}{dz} & =  & - \frac{2\, \dot{Q}_d}{3 k_B (1+z) H}
\end{eqnarray}
These two equations represent the change in temperature of the baryons and DM due to baryon-DM interactions. For this case, $f_{\chi,{\rm int}}$ dependence of these equations are shown in reference \citep{Venumadhav:2018uwn, Munoz:2015bk}.
\subsection{Decay/annihilation of DM particles}
\label{sec-2-sub-2.4}
\begin{figure}
	\centering
	\includegraphics[width=0.7\linewidth]{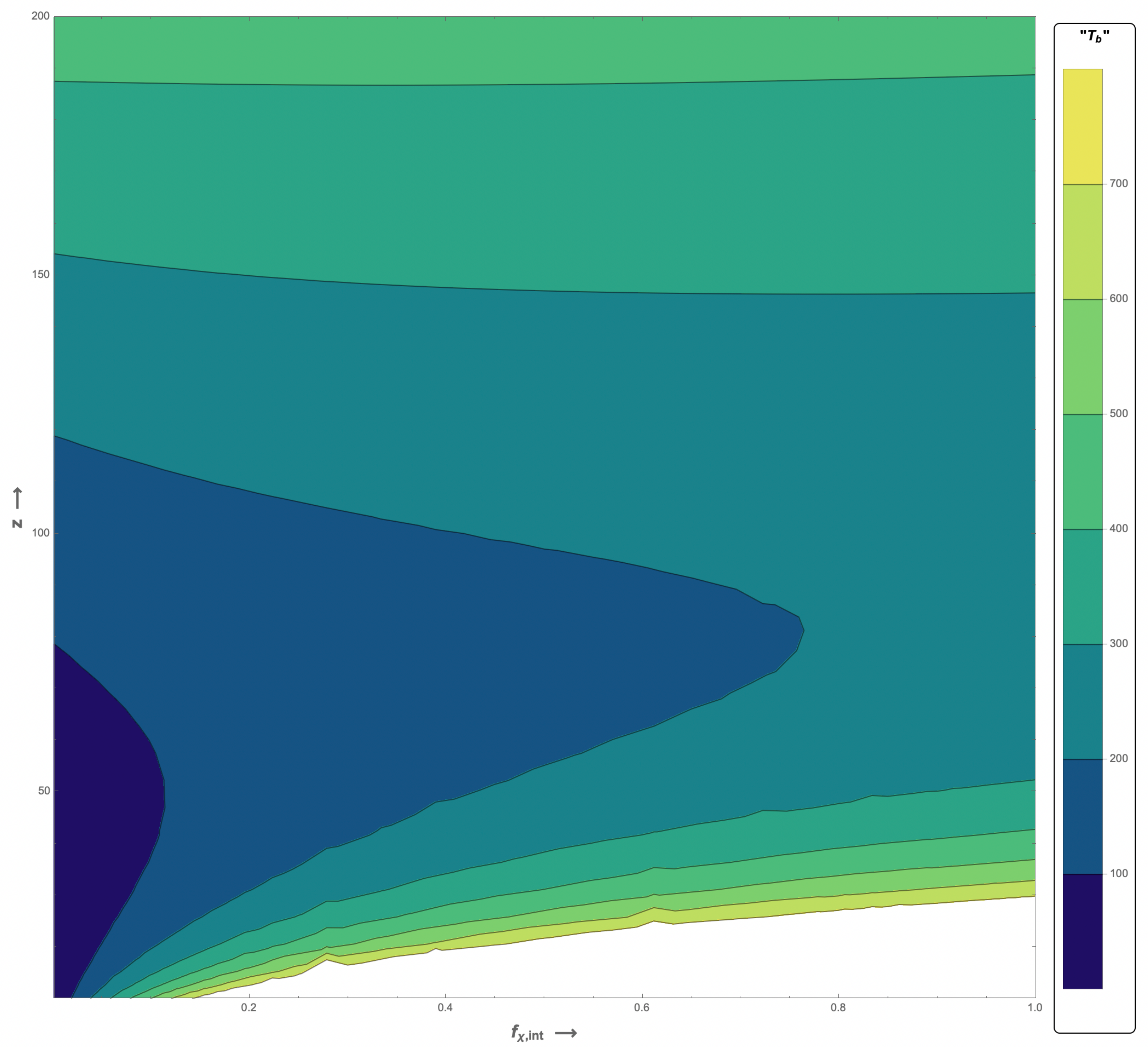}
	\caption{\textbf{BDM+$B_0$+Decay:} In this case, we have considered following parameters: $B_0=0.003$ nG, $n_B=-2.9$, $\lambda_B=0.25$ Mpc, $m_\chi=0.1$ GeV, $\sigma_0=3\times 10^{-45}~\text{m}^2$, $\tau=2\times 10^{25}$ sec, $f_{\chi,{\rm eff}}=0.6$, $f_{\chi,\text{dec}}=0.6$. In this plot, we have only considered redshift $z=10-200$. At higher redshift, we have found that there are no visible effect on the baryon temperature. }
	\label{fig:decay-z-fxint}
\end{figure}
It is known that both stable and unstable standard model particles and their antiparticles are found in the exotic decay of particles. The unstable products eventually decay into stable particles, such as electrons, photons, protons, and neutrinos. These particles and their antiparticles lose energy either adiabatically or interact with the particles before or after the recombination epoch and transfer energy to the photons and light element atoms (for example, electrons and nuclei). In the present work, we will limit ourselves to the injection of electron-positron or photons from the decay of DM ($\chi \rightarrow e^+ e^-$ or $\chi \rightarrow \gamma \gamma $) and from the annihilation of the DM ($\chi \chi\rightarrow e^+ e^-$ or $\chi \chi\rightarrow \gamma \gamma$) with a monochromatic energy spectrum (for the other possible decay channels, see reference \cite{Slatyer:2015jla}). We have ignored the energy deposition from the processes where protons/antiprotons and neutrinos are decay products. Since neutrinos are invisible to the medium because of their tiny mass and interaction cross-section, they carry a small fraction of the total output energy. Similarly, protons and antiprotons inject very small energy into the medium. Hence we have ignored the decay channels, where protons/antiprotons are main decay products \cite{Weniger:2013hja, Slatyer:2016rj, Liu:2019bbm, Cang:2020exa}. The $\gamma\gamma$ final state of the decay of sterile neutrinos from the decay mode $\gamma \nu$ is relevant to the unstable gravitinos with mass below the $W$ gauge bosons \cite{Liu:2018uzy}. It is thus clear that the chosen channels are mostly dictated by simplicity since, in most decay models, a continuous spectrum of standard model particles is emitted. Our formalism in the present work applies to any final state with a generic $e^\pm$ and photons $\gamma$. Hence the results obtained are mostly indicative and do not represent the actual, more complicated energy injected process from the decay/annihilation of the DM particles.

The additional particles and photons injection have three main contributions: i). the number of non-thermal particles grows; however, their average energy density decreases due to the interaction with the relic photons, ii). when the energy density of the non-thermal particles reaches keV scale, they start interacting with the atoms (mostly hydrogen atoms) and induce Lyman-$\alpha$ excitation of the hydrogen, which in results a higher number of electrons in $n=2$ states than the ground state of hydrogen and hence increased ionization fraction \cite{Galli:2013dna}, and iii). can directly heat the gas. The first two effects change the ionization fraction, and the last terms lead to a change in the baryon temperature. It has been shown that the injected energy is often not absorbed \textit{on-the-spot}; instead, it is redshift away before being deposited \cite{Galli:2009zc, Slatyer:2016rj}. Therefore, the annihilation or decay of the DM into the standard model (SM) particles inject energy (denoted by, $\Gamma_{\rm dep}=\frac{dE}{dV\, dt}|_{\rm dep}$) into the universe and leads to an additional ionization (`$i$'), excitation (`$\alpha$') and heating (`$h$') of the medium. The most critical parameters that affect the energy injection from the decay/annihilation are, a). decay/annihilation channels of the DM particles, b). mass of the DM particles, and c). the redshift at which the energy injections occur. The injected energy is parameterized in terms of the redshift as
\begin{equation}
	\Gamma^{\rm dep}_c= f_c(z) \, \frac{dE}{dV\, dt}\bigg|_{\rm inj}.
\end{equation}
Here the dimensionless function, $f_c(z)$ represents the energy deposition efficiency in the medium from a relevant channels \cite{Shull:1985jm, Chen:2003gz, Galli:2009zc, Finkbeiner:2012prl, Slatyer:2012yq, Poulin:2015pna, Slatyer:2015kla, Slatyer:2015jla,  Slatyer:2016rj, Slatyer:2016qyl}. The subscript `$c$' represents three different channels and is given by `$c=\{i, \alpha, h\}$'. This function encodes all the physics of the energy deposition accounted for all additional standard model electromagnetic processes due to the annihilation/decay in IGM \footnote{Usually this function $f_c$ depends on not only $z$ but also on ionization fraction $x_e$, i.e, $f_c(z, x_e)$. However, for energy injection $\geq 10$ MeV, $f_c(z, x_e)\simeq g(z) \chi_c(x_e)$ is a good approximation. This correspondence to a factorization between the high energy process (which determine the function $g(z)$) and low energy processes, responsible for the absorption (determined by function $\chi_c(x_e)$) \cite{Galli:2013dna, Slatyer:2015kla}.}.  In references \cite{Finkbeiner:2012prl, Giesen:2012rp}, it has been shown that, up to first-order approximation, redshift dependence of the function $f_c(z)$ can be ignored, as current CMB data are sensitive to energy injection over a relatively narrow range of redshift ($z\sim 1000-600$) \cite{Mohanty:2021na}. Therefore, $f_c(z)$ can be replaced by a constant number $f_{\rm eff}$ and for sake of simplicity, we have approximated $f_c(z)=f_{\rm eff}$ a constant efficiency factor over the entire redshift range. This assumption means that the injected energy is instantly deposited in the intergalactic medium (IGM). Normally the $f_{\rm eff}$ depends on the mass of DM and the EM branching ratio, i.e., decay channels. As the DM annihilation or decay into SM particles injects energy into the medium, an additional ionization, excitation, and heating occur. The rate of energy injection from these processes is given by
\begin{equation}
	\label{eq:dec-anni1}
	\frac{dE}{dV dt}\bigg|_{\rm inj}= 
	\begin{cases}
		(1+z)^3 \frac{f_{\chi, {\rm dec}}}{f_{\chi_{,\rm in}}^{-1}} \Omega_{\rm DM} c^2 \rho_c \frac{Exp(-\frac{t}{\tau})}{\tau}, & \text{for decay}  \\
		(1+z)^6 \left(\frac{f_{\chi, {\rm ann}}^2}{f_{\chi,{\rm in}}^{-1}}\right) \Omega_{\rm DM}^2 c^2 \rho_c^2 \frac{\langle\sigma v\rangle }{m_\chi}, & \text{for annihilation}
	\end{cases}
\end{equation}
where $f_{\chi, {\rm dec}}$ and $f_{\chi, {\rm ann}}$ are the decay/annihilation fractions by mass of the DM particles, $\Omega_{\rm DM}$ is the present-day total abundance of cold DM, $\rho_c$ is the critical density of the universe, $\tau$ is the lifetime of the decaying particle (related to the decay rate $\Gamma_d$ by $\tau=\Gamma_d^{-1}$). $m_\chi$ is the DM particle mass, and $\langle \sigma v\rangle$ is the thermally averaged self-annihilation cross-section. If annihilation occurs between two indistinct particles, an additional factor of $1/2$ for Majorana fermions and $1/4$ for the Dirac fermions should be multiplied. In the present work, we only consider that annihilating particle is indistinguishable. $f_\chi=1$ represents the complete decay/annihilation of the DM particles. For the DM annihilation, we have defined a new more condense efficiency parameter  $p_{\rm ann}=f_{\rm eff} \langle \sigma v \rangle/ m_\chi$. It is important to note that we have not considered the structure formation. In reference \cite{Slatyer:2012yq} authors have shown that at redshift $z\leq 30$ there might be measurable effects on the 21-cm line of hydrogen from the structure formation \cite{Furlanetto:2006wp, valdes:2007fmr}. In this case, halo contributes to additional factors in equations \eqref{eq:dec-anni1} as erfc$\left(\frac{1+z}{1+z_h}\right)\bar{f}_h$, where $z_h$ is the redshift at which halo is formed and $\bar{f}_h=\frac{200}{3}(1+z_h)^3 f_{\rm NFW}$ is the normalization factor obtained from N-body structure formation simulations \cite{Giesen:2012rp}. Change in the temperature of the universe due to the energy deposition from the DM decay/annihilation is given by 
\begin{eqnarray}
	\frac{dT_{b_{\rm dec/ann}}}{dz}  = -\frac{1}{H(z) (1+z)}\frac{2 f_h(z)}{3 k_B n_H (1+f_{He}+x_e)} \frac{dE}{dV dt}\bigg|_{\rm inj}.
	\label{eq:Tbdecann-1}
\end{eqnarray}
The heating term on the right-hand side of the equation (\ref{eq:Tbdecann-1}) comes when a non-zero DM decay/annihilation and function $f_h(z)$ is the dimensionless efficiency function which parameterizes the extra heat injected to increase the gas temperature. The contribution to the ionization fraction from the DM decay/annihilation is given by
\begin{eqnarray}
	\frac{d x_{e}^{\rm dec/ann}}{d z} =-\frac{1}{(1+z)H(z)}I_X(z).	\label{eq:ion1}
\end{eqnarray}
The term on the right hand side of equation (\ref{eq:ion1}) represents the ionization rate due to extra injection of energetic particles and it can be split into two terms as: $I_X=I_{X_i}+I_{X_\alpha}$ \cite{Chen:2003gz, Poulin:2016anj, Slatyer:2016qyl, Chluba:2010jc}.  Here $I_{X_i}$ and  $I_{X_\alpha}$ represent the hydrogen ionization from the ground state ($n=0$) and by additional Ly-$\alpha$ photons, which boosts the population at the $n=2$ state and hence increases the photoionization by CMB photons. These two terms $I_{X_i}$ and $I_{X_\alpha}$ are defined as
\begin{eqnarray}
	I_{X_i} & = & -\frac{1}{n_H(z)E_i} \frac{d E}{dV \, dt}\bigg|_{{\rm dep}, i}\\
	I_{X_\alpha} & = & -\frac{1-C}{n_H(z)E_\alpha} \frac{d E}{dV \, dt}\bigg|_{{\rm dep}, \alpha}
\end{eqnarray}
The two constants $E_i=13.6$ eV and $E_\alpha=E_{21}=10.2$ eV are the average ionization per hydrogen atom and the Ly-$\alpha$ energy respectively. Parameter $\mathcal{C}$ is known as the inhibition factor, and it represents the number of excitation events that contributes to the ionization (each event deposits energy equals 10.2 eV). For a direct ionization $\mathcal{C}=0$. In the present work, we neglect the extra heat injected from the Helium ionization as the Helium ionization is sub-dominant \cite{Galli:2013dna, Slatyer:2015kla} and hence it will not affect the results significantly. Therefore, the contribution of the ionization and excitation is given by
\begin{equation}
		\frac{d x_{e}^{\rm dec/ann}}{d z} =  \frac{1}{H(z)(1+z)}\left[\frac{f_i(z)}{E_i n_H}+\frac{(1-\mathcal{C}) f_\alpha(z)}{E_\alpha n_H}\right]\frac{dE}{dV dt}\bigg|_{\rm inj},
	\label{eq:xedecann-1}
\end{equation}
here $f_i(z)$ and $f_\alpha(z)$ are related to the ionization and excitation efficiency factors respectively. Thus, energy deposition from the DM annihilation/decay will increase the ionization fraction and IGM temperature after the recombination epoch. Another important effect that can also affect the ionization and thermal history is DM clustering after the formation of halos. In the cold DM scenario, DM clusters into halos due to gravity at various ranges of masses; for example: of size $10^{-11}-10^{-3}$ M$_\odot$ for WIMPs to DM clusters halos of size $10^{15}$ M$_\odot$ \cite{Bringmann:2009vf}. It has been shown that the annihilation rate is boosted late due to DM halos and the DM decay maintains a steady energy injection rate \cite{Cang:2020exa}. In spite of this we have confined our study only to redshift $z\simeq 10$ and avoided the effects of the DM halo formation.
\subsection{Complete set of equations}
\label{sec-2-sub-2.5}
In the previous subsections (\ref{sec-2-sub-2.1}-\ref{sec-2-sub-2.4}), we have first discussed standard thermal and ionization history of the universe. Later, we presented a brief discussion on the heating of the gas temperature from the magnetic heating and DM annihilation/decay along with the energy exchange between the baryons and the DM particles due to a non-standard interaction between baryons the DM particles. After considering all these effects, the baryon temperature is $T_b=T_{b}^{\rm sta}+T_b^{\rm cool}+T_b^\delta+T_{b}^{\rm mag}+T_{b}^{\rm BDM}+T_{b}^{\rm dec/ann}$, and $x_e=x_{e}^{\rm sta}+x_{e}^{\rm dec/ann}$. Therefore, the final equations, we have used following equations to study the thermal and ionization history of the universe
\begin{eqnarray}
	\frac{dT_b}{dz} & = & \frac{dT_b^{\rm sta}}{dz}+\frac{dT_b^{\rm cool}}{dz}+ \frac{dT_b^\delta}{dz}+ \frac{dT_b^{\rm mag}}{dz} \nonumber \\
	& +& \frac{dT_b^{\rm BDM}}{dz} +\frac{dT_b^{\rm dec/ann}}{dz}, \label{eq:baryon_temp-2}  \\
	\frac{dT_{d}}{dz}& = &\frac{2T_{d}}{(1+z)} - \frac{1}{1.5 k_B n_H\, (1+z)H}\frac{d{Q_{d}}}{dt}\, ,
	\label{eq:dm_temp-1} \\
	\frac{d\text{v}}{dz}& =&  \frac{\text{v}}{(1+z)} + \frac{D(\text{v})}{(1+z)H} \, ,
	\label{eq:vel-1}
\end{eqnarray}
In equation (\ref{eq:baryon_temp-2}), first and second term on the right hand side represents the effects of standard Compton scattering and cooling from the bremsstrahlung, collisional excitation, recombination and collisional ionization cooling. The third term is solely coming from the density perturbation created by the magnetic fields.  The fourth, fifth and sixth terms in equation \eqref{eq:baryon_temp-2}, represents the terms related to the magnetic heating, BDM interaction and the decay/annihilation respectively. The ionization equation under the present study is given by
\begin{eqnarray}
	\label{eq:ion2}
	\frac{d x_e}{d z} = \frac{d x_e^{\rm sta}}{d z}+ \frac{d x_{e}^{\rm dec/ann}}{d z}
\end{eqnarray}
Here the two terms on the right hand side are given in equations \eqref{eq:ioni1} and \eqref{eq:xedecann-1}. In our numerical simulation, we have used equations \eqref{eq:baryon_temp-2}-\eqref{eq:vel-1}, along with the magnetic energy density evolution equation \eqref{eq:magen-1} and  the ionization fraction evolution equation \eqref{eq:ion2} to understand the thermal and ionization history of our universe in IGM. In our work, $B_0$, $n_B$, $\lambda_B$, $\sigma_0$, $m_\chi$, $\langle \sigma v\rangle/m_\chi$ and $\tau$ are used as a parameter. In different cases, different parameter combinations have been used to solve these equations.
\section{Result and Discussion}
\label{sec-3}
In the present work, our primary motivation is to understand the thermal and ionization history of the IGM between redshift $1100-10$. 
To avoid the effect of structure formation, we limit our study upto redshift $z\sim 10$. We have computed the evolution of temperature, ionization fraction, and $y-$parameter of the tSZ effect to understand the processes involved. We have included the following effects in addition to the standard terms related to Thomson Scattering and the adiabatic cooling: 
\begin{itemize}
	\item cooling effects like bremsstrahlung, collisional excitation, recombination and collisional ionization cooling,
	\item magnetic decay from the ambipolar diffusion and turbulent process,
	\item effects from the interaction of baryons to the DM particles (BDM processes), 
	\item processes involving the decay or annihilation of the DM particles. 
\end{itemize}
In particular, the BDM interaction with interaction cross-section of the form $\sigma=\sigma_0 ({\rm v}/c)^{n}$ and decay/annihilation channels $\chi\rightarrow (e^+e^-,  ~\text{or} ~ \gamma\gamma)$,  $\chi \chi\rightarrow (~e^+ e^-, ~\text{or} ~\gamma \gamma)$ are considered in the present work.  Important thing to be noted here is that the energy deposition efficiency function $f_c(z)$ is constant over the redshift of interest. The assumption is made to simplify our numerical simulations. Here $B_0$, $n_B$, $\sigma_0$, $n$, $f_{\chi,{\rm int}}$, $f_{\chi,{\rm ann}}$, $f_{\chi,{\rm dec}}$, $p_{\rm ann}=f_{\rm eff}\langle \sigma v\rangle/m_d$ and $\tau$ are considered as a parameter. For our convenience, following values are used $B_0=0.003$, $1.0$ nG, $n_B=-2.9$ and $-1.0$, $n=-4$ and $\lambda_B=0.25, 1.0$ Mpc. In the upcoming subsections, we briefly explain the role of various effects on the temperature of baryons and DM and on the ionization fraction. In subsection \eqref{sec-3-the-ion}, the temperature and ionization fraction of the baryons along with the DM temperature evolution is presented. In this subsection, we also discuss the role of partial decay/annihilation on the temperature of the baryons. Subsection \eqref{sec-3-tsz-effect} contains the impact of various effects on the baryon temperature and hence the CMB anisotropy during redshift 1100-10. Finally, subsection \eqref{sec-3-DP-B0} holds a discussion on the bound on the magnetic fields from the decay/annihilation of the DM particles in our present context.
\subsection{Thermal and ionization history} \label{sec-3-the-ion}
In this subsection, we briefly discuss the evolution of baryon, DM temperature, and the ionization fraction from redshift $z=1100-10$ in various scenarios. Below in all figures, black dashed and solid black lines represent the CMB temperature and baryon temperature in the standard scenario respectively. Based on the effects considered, this subsection can be separated into the following cases:
\begin{enumerate}
	\item \textbf{BDM, $B_0$:} In this scenario, the effects of the magnetic field and the BDM interaction are only considered (see figures \eqref{fig:BDM-B0-net} and \eqref{fig:BDM+B0+fxint}). Solid and dashed (except the black dashed line) lines in this plot represent the evolution of baryon and dark matter temperature, respectively. In reference \cite{Pandey:2020hfc}, authors studied this scenario at a great length. Some of the most relevant scenarios for the current project have been discussed here. In figure \eqref{fig:BDM-B0-net}, we fix $f_{\chi,{\rm int}}=1.0$, $\sigma_0 =3.0\times 10^{-45}$ m$^2$ and $m_\chi=1.0$ GeV and compare the results for the BDM only case to the results of the case when BDM+$B_0$ are present. We now show the role of $f_{\chi,{\rm int}}$ in determining the baryon and DM temperatures in figure \eqref{fig:BDM+B0+fxint}. In figure \eqref{tem-evolution-BDM-B0-only}, the solid red line corresponding to $B_0=0.003$ nG and $n_B=-2.9$ (nearly scale-invariant spectrum) and the solid green line (only BDM) show almost identical behavior except for lower redshift where minor effects of magnetic fields are seen. However, for $B_0=0.003$ nG and $n_B=-1.0$, baryon temperature follows the standard pattern of the baryons until redshift $z \approx 30$, at which point it turns and later crosses the CMB temperature. We have observed that the baryon temperature for $B_0 \geq 1$ nG and $n_B=-2.9$ becomes more prominent at a very early stage, i.e., the redshift at $z\simeq 800$. Even starting with the temperature $T_d\approx 0$ K, we can see that the temperature of the DM starts to increase as we move to a smaller redshift. The BDM interaction caused the energy transfer from the baryon to the DM. It is important to note that magnetic fields do not directly affect DM temperature but rather magnetic fields through BDM interactions. Figure \eqref{xe-evolutio-BDM-B0-only} contains the evolution of the ionization fraction of the neutral hydrogen for various combinations of parameters. In this case, we have compared the ionization fraction in the presence of magnetic fields and BDM interaction to the standard scenario. It is visible from the plot that the larger is the value of $B_0$, the larger is the ionization fraction. Starting from fully ionized state at $z\approx 1100$, $x_e$ decreases rapidly and attain $x_e\approx 10^{-4}$. However, extra energy injected from the decaying magnetic field leads to a larger value of $x_e$ at a lower redshift. In these plots, we have considered $f_{\chi,{\rm int}}=1$, which means all DM particles are interacting with the baryons. This nature changes when we assume a fraction of the total DM particles are interacting with the baryons. This can be seen in figure \eqref{fig:BDM+B0+fxint}. In the left panel of this figure, (i.e. figure \eqref{BDM-B0-Tbfxint-z-Stan-B0003-BDM-sig-45-mdm0.1GeV}), we set $B_0=0.003$ nG, $\lambda_B=0.25$ Mpc, $\sigma_0=3\times 10^{-45}$ m$^2$, and DM mass $m_\chi=0.1$ GeV. It is visible that the larger the value of $f_{\chi,{\rm int}}$, the more significant will be the impact on the baryon temperature. The explanation for this trend is that many DM particle interactions lead to more energy extraction from baryons. In the case of BDM+$B_0$, we can observe that the influence is stronger for bigger values of the $f_{\chi,{\rm int}}$ at lower redshift \eqref{BDM-B0-Tbz-fxint-Stan-B003-BDM-sig-45-mdm0.1GeV}. At redshift $z=100 (\text{and}~50)$, baryon temperature drops from $T_b\approx 100 (\text{and}~40)$ K to $10$ K and $2$ K respectively. However, at high redshift, no significant effect of $f_{\chi,{\rm int}}$ is found. A similar trend is noted in the DM temperature.
\item \textbf{BDM, $B_0$, Annihilation:} 
In this case, we have chosen three cases: one when only DM particles are annihilating (no BDM, no $B_0$), the second when we include magnetic field effects in addition to DM annihilation, and finally when we use BDM interaction+effects of magnetic fields+DM decay/annihilation. Following annihilation channels are of our interest: $\chi\chi\rightarrow \gamma\gamma~\text{or}~ e^+ e^- $. The baryon and DM temperature evolution is explored in two scenarios in figure \eqref{fig:B-DM-Int-1}: when just DM annihilation is considered and when DM annihilation+BDM interaction is present. In figure \eqref{fig:B-DM-Int}, we add the effects of the magnetic fields and the DM annihilation. Figure \eqref{fig:detail-fxint-sig} shows the effects of BDM interaction+decaying magnetic fields+DM annihilation. In this figure, we show $f_{\chi,{\rm int}}$ dependence of the baryon-DM temperature. In figure \eqref{fig:baryon-T-B003-1}, we plot the baryon and DM temperature for various combinations of the parameter and the CMB temperature together. In this case, we have considered $f_{\chi,{\rm int}}=1.0$. Clearly, with the introduction of DM annihilation, additional energy injection leads to a larger baryon temperature than the standard scenario. Baryon temperate follows each other for the most part for a particular annihilation channel. This behavior is also reflected in the plots of the DM temperature (in figure \eqref{fig:baryon-T-B003-1}) and the ionization fraction (in figure \eqref{fig:baryon-T-sig-42-1}). When the effect of the BDM interaction is incorporated, things start to change. All other lines show a small effect from DM annihilation and BDM+DM annihilation, with the exception of the annihilation channel $\chi\chi \rightarrow \gamma\gamma$ with DM mass $m_\chi=0.1$ GeV (darker green solid line). The darker green lines corresponding to BDM+annihilation channel $\chi\chi\rightarrow \gamma\gamma$ with DM mass $m_\chi=0.1$ GeV show a drastic change in the behavior at low redshift due to the fact that small mass DM has smaller annihilation parameter $\langle \sigma {\rm v}\rangle/m_\chi$, and hence lower energy injection from the decay of DM particles (check the figure-5 of reference \cite{Liu:2018uzy}. However, BDM interaction, in this case, will extract energy and lower the baryon temperature further. 
Things get changed when we include the effect of magnetic fields in addition to the DM annihilation (see figures \eqref{fig:B-DM-Int} and \eqref{fig:detail-fxint-sig}). In the presence of DM annihilation, the the heating from the magnetic fields leads to higher baryon temperature and ionization fraction at a given redshift than when solely DM annihilates. It's also worth noting that the ionization fraction increases with the mass of the annihilating DM particles. Figure \eqref{fig:detail-fxint-sig} shows the dependence of the baryon and DM temperature on $f_{\chi,{\rm int}}$. In figure \eqref{fig:Tb-Td-max-min-fxint-plot-arun}, there are two regions, one light green and another blue-shaded one. The area between solid red ($f_{\chi,{\rm int}}=0.001$) and solid blue ($f_{\chi,{\rm int}}=1.0$) represents the region over which baryon temperature varies with respect to $f_{\chi,{\rm int}}$ at a given redshift. Similar is the behavior of the DM temperature, shown in the red-dashed line (for which $f_{\chi,{\rm int}}=0.001$) and blue-dashed line (for which $f_{\chi,{\rm int}}=1.0$) (see figure \eqref{fig:Tb-Td-max-min-fxint-plot-arun}). In figure \eqref{fig:TB-Tdsig-vs-fxint}, $f_{\chi,{\rm int}}$ dependence of the baryon and DM temperate are shown. Clearly, there is no visible role of $f_{\chi,{\rm int}}$ on the baryon or DM temperature at redshifts of $z=1000$ or even for $z=500$. At lower redshifts, however, the parameter $f_{\chi,{\rm int}}$ has a stronger influence on both baryon and DM temperature. Baryon temperature  $T_b(z=100)\approx 150$ K and $T_b(z=50)=45$ K at $f_{\chi,{\rm int}}=0.001$,  drops to $T_b(z=100)=50$ K and $T_b(z=50)=3.9$ K for $f_{\chi,{\rm int}}=1.0$. Contour plots for the baryon temperature in $\sigma_0-f_{\chi,{\rm int}}$ at redshift $z=20$ plane and $z-f_{\chi,{\rm int}}$ plane in figures \eqref{fig:Tb-Td-vs-fxint} and \eqref{fig:Tb-vs-fxint-contour} respectively. Even at smaller values of $f_{\chi,{\rm int}}$ but for a large values of $\sigma_0$, baryons temperature at redshift $z\approx 20$, have larger values compared to the case when $\sigma_0$ is small but large values of $f_{\chi,{\rm int}}$ (see the figure \eqref{fig:Tb-Td-vs-fxint}). The in-set figure, in figure \eqref{fig:Tb-Td-vs-fxint} represent the contour plot of $T_b$ for a smaller range of $\sigma_0$ and $f_{\chi,{\rm int}}$. Clearly when BDM interaction and magnetic fields are present, larger effects happens on baryon temperature in presence DM annihilation at small values of $f_{\chi,{\rm int}}$ and larger values of $\sigma_0$. For a DM of mass $m_\chi=0.1$ GeV, CMB bound on the BDM interaction cross-section $\sigma_0 \approx 3.0\times 10^{-45}~\text{m}^2$  gives a baryon temperature $T_b\approx 3.0$ K at $f_{\chi,{\rm int}}\approx 0.1$ and $T_b\approx 0.5$ K at $f_{\chi,{\rm int}}\approx 0.5$. It is also evident that the at lower redshift, evolution of $T_b$ is highly dependent on the values of $f_{\chi,{\rm int}}$ (see figure \eqref{fig:Tb-vs-fxint-contour}).
\item \textbf{BDM, $B_0$, Decay:} 
In the case of DM decay, we have considered following decay channels: $\chi\rightarrow e^- e^+~\text{or}~ \gamma\gamma$. Again we have subdivided this scenario: (i). Only DM decay, (ii). Decay+$B_0$, (iii). Decay+BDM and (iv). Decay+$B_0$+BDM. In figures \eqref{fig:detail-1}-\eqref{fig:detail-2}, baryon and DM temperature along with the ionization fractions are shown in the above mentioned scenarios. Figure \eqref{fig:decay-z-fxint} shows the contour plot of the baryon temperature in $z-f_{\chi,{\rm int}}$ plain. The effects of DM decay on baryon temperature (and ionization fraction) are compared in figures \eqref{fig:Td-B05-1} (and \eqref{fig:Td-sig042-1}) respectively to scenarios where only BDM interaction is present and to standard baryon temperature (and ionization fraction) evolution. Figure \eqref{fig:detail} shows the evolution of baryon temperature and the ionization fraction when DM decay and the magnetic fields are present. In figure \eqref{fig:Td-B05-1}, it is clear that the DM decay injects additional energy, which in turn enhances the baryon temperature. Up to $z\approx 800$, the temperature of baryons follows that of the CMB but subsequently exceeds it at lower redshifts. It is also worth noting that, regardless of the decay channels, the temperature of baryons remains higher than that of the CMB for redshift $z<800$ and even can reach $4000$ K at redshift 10. Figure \eqref{fig:Td-sig042-1} shows that the ionization fraction declines at first, then rises to a nearly fully ionized state, starting from a fully ionized state at $z \approx 1100$. When the effects of magnetic fields are turned on, again, the baryon temperature remains higher than the CMB temperature for the majority of the time for selected magnetic field settings (see figure \eqref{fig:Td-B05}). The red dashed lines show the baryon temperature evolution when only magnetic fields are considered (see figure \eqref{fig:Td-B05}). Except for $z<17$, when we turn on the DM decay together with $B 0$, the influence of magnetic fields in the absence of DM decay (red dashed line) dominates most of the time.

Similarly, in figure \eqref{fig:Td-sig042}, the ionization fraction is always higher in case of decay of the DM particles. It is to be noted here that the universe started with a fully ionized state at redshift $z=1100$, and during the dark ages, it degrees drastically. However, due to the additional energy injection from the decay of the DM particles, it begins to reionize at lower redshift $z$ and becomes fully ionized at redshift $z\leq 10$. Because the effect of structure formation has not yet been factored into the present work, it's interesting to examine the behavior of baryon temperature and ionization fraction in the IGM when DM particles are decaying or annihilating. This has been left for our future work. We have compared the baryon temperature and ionization fraction in the case of DM decay with the case of BDM+DM decay in figure \eqref{fig:decBDM-9}. In figure \eqref{fig:decBDM-tbtd}, the solid and dashed lines indicate the evolution of the baryon temperature and DM temperature with respect to redshift. As discussed above, the BDM interaction extracts heat, and hence in the absence of DM decay, baryon temperature remains lower than $T_{b,{\rm std}}$ (solid blue line). Except in the redshift region $20<z<800$, baryon temperature remains the same in the presence of BDM interaction (solid sky-blue line) and the absence of BDM interaction (solid red line) for decay channel $\chi \rightarrow e^- e^+$. In this case, BDM interaction extracts the heat and lowers the baryon temperature. Because both the solid red and solid sky-blue lines merge at redshift $z<20$, it suggests that BDM interaction has little effect at lower redshifts for the chosen set of parameters. As can be observed in plot \eqref{fig:decBDM-tbtd}, these characteristics are the same for all other channels. The evolution of the ionization fraction for the given set of parameters is depicted in figure \eqref{fig:decBDM-xe-9} in two cases: when the sole influence of DM decay is evaluated and when both DM decay and BDM interaction are considered. The discussion from plot \eqref{fig:Td-sig042} applies to this plot as well. In figure \eqref{fig:detail-2}, dependence of baryon temperature on $f_{\chi,{\rm int}}$ is compared from BDM$+B_0+$Decay to the standard scenario. Baryon temperature and CMB temperature exhibit a crossover for two channels $\chi \rightarrow e^- e^+$ and $\gamma \gamma$ at redshifts $z \approx 80$ and $250$, respectively, in the left panel of this figure (see figure \eqref{fig:dec-tbtd}. Here we have fixed the value of $f_{\chi,{\rm int}}$. The dependence of baryon temperature on $f_{\chi,{\rm int}}$ at a given redshift is shown in figure \eqref{fig:dec-tbtd-fxint}. Clearly baryon and DM temperature have similar dependence on $f_{\chi,{\rm int}}$. At a large redshift, irrespective of decay channels, there is no visible dependence of the baryon and DM temperature on the $f_{\chi,{\rm int}}$ similar to the annihilation case. As we go from large values of $f_{\chi,{\rm int}}$ to smaller values, baryon and DM temperature decreases. However, in the case of annihilation, we have seen that the behavior is the opposite. The temperature of both baryons and DM rises at first, then settles at a particular value at lower values of $f_{\chi,{\rm int}}$ (see figure \eqref{fig:Tb-Td-max-min-fxint-plot-arun}). A contour plot of baryon temperature in a redshift $z-f_{\chi,{\rm int}}$ plane is depicted in figure \eqref{fig:decay-z-fxint} for a particular set of parameters (the values of parameters are given in caption). Several color zones represent the temperature of baryons. The temperature of baryons is more dependent at lower redshifts and larger values of $f_{\chi,{\rm int}}$. The baryon temperature has the least influence on $f_{\chi,{\rm int}}$ in the redshift range $\sim 10-100$. This range is exciting when considering the effects of structure formation or DM haloes.
\end{enumerate}
\subsection{Thermal SZ effect induced by the decay/annihilation of DM and the y-parameter} 
\label{sec-3-tsz-effect}
\begin{figure}
\centering
	\includegraphics[width=0.8\linewidth, keepaspectratio]{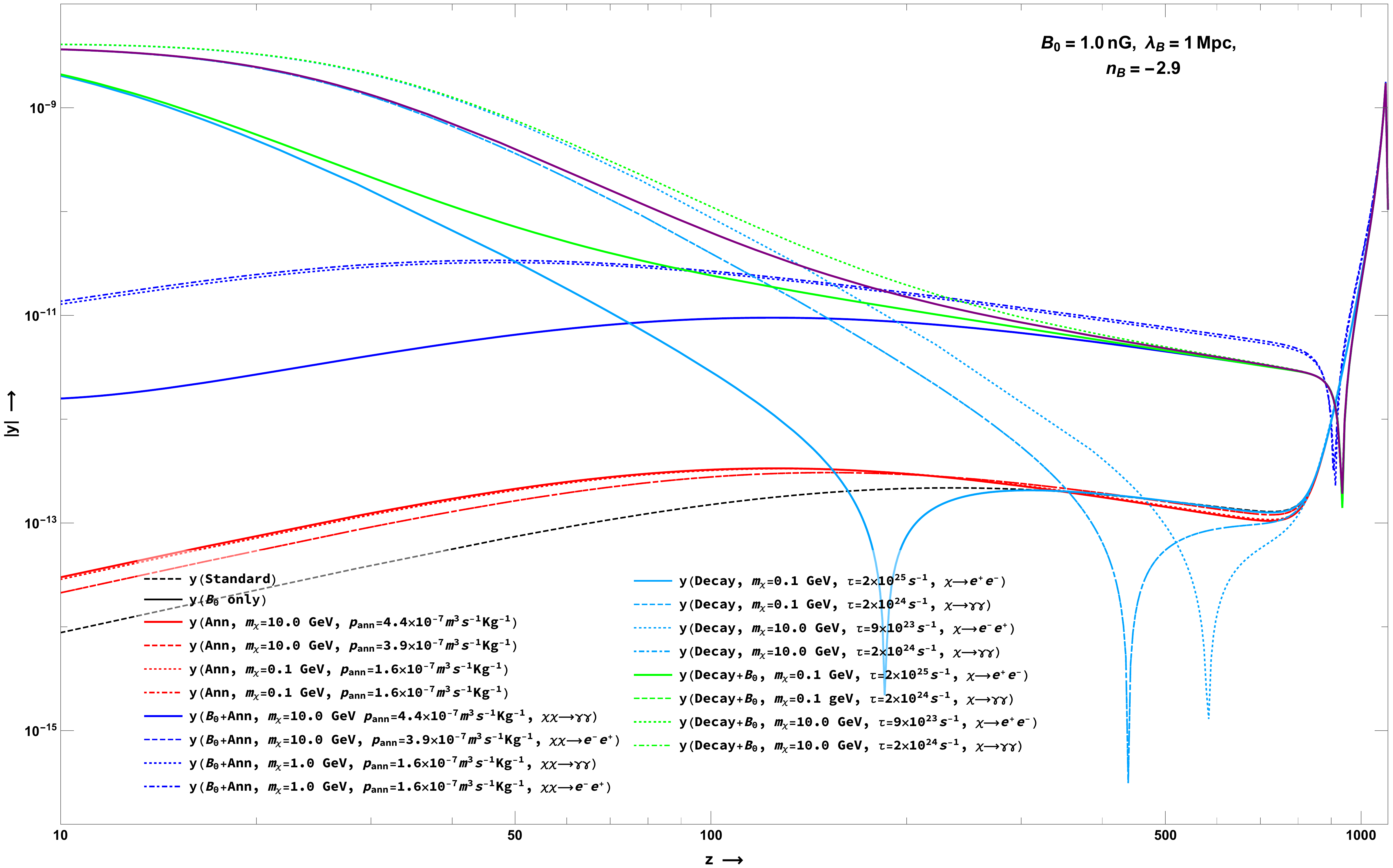}
	\caption{Baryon and Dark matter temperature in presence of baryon dark matter interaction and magnetic fields but no decay/annihilation processes. Here we assume one to one interaction is between the baryons and dark matter particles (i.e. $f_{\chi_{,\rm in}}=1$.). Also coherence scale of the magnetic fields are $\lambda_B= 0.25$ Mpc. Color of the lines are explained in the legends. This plot is similar to the result obtained in reference.}
	\label{fig:tb-sz-1}
\end{figure}

In this section, we discuss the tSZ effect for various possible scenarios. The $y-$ parameter is a quantity that gives information on the distortion in the CMB spectrum. We compare the $y-$ parameter for the standard scenario with the case when we include the effects of magnetic fields, BDM interactions, and decaying magnetic fields via ambipolar diffusion and turbulent decay. We also include the effect of background density perturbations from the magnetic fields and the cooling via Bremsstrahlung, collisional excitation, recombination, and collisional ionization cooling. The $y-$parameter on a sky plane in direction $\hat{n}$ is defined as
\begin{equation}
	y(\hat{n})=\int dz \frac{k_B \sigma_T}{m_e c} \frac{\left[x_e n_H (T_b-T_{\rm CMB})\right]|_{\hat{n},z}}{(1+z)H(z)}.
	\label{eq:tsz-formula}
\end{equation}
Since we are mainly interested in studying the background evolution, we have only the average quantity in the sky plane. From the equation \eqref{eq:tsz-formula}, we can see that the $y-$parameter not only depends on ionization fraction and background baryon number density, but it depends on the difference of baryon temperature with the CMB temperature at a given redshift. In the previous subsection of the result and discussion section, we have seen that the decay/annihilation could have significantly affected at low redshift. What is important here is that the behavior of the value of $y-$ parameter depends on the difference between baryon temperature and the CMB temperature at a given redshift in a particular direction of the sky plane. Therefore, given the sign of this difference, the $y-$parameter changes its sign from positive to negative and vice versa. In figure \eqref{fig:tb-sz-1}, we have shown the evolution of absolute values of $y-$parameter with respect to $z$. In this plot, we have given a color code for each case. 
%
In this plot, red-colored lines denote the case when we have only considered annihilation. The blue color lines represent the annihilation of the DM particles in the presence of the background magnetic field. The decay case is shown in light blue (only decay) and green colors (with decay and background magnetic fields). When background magnetic fields are present, we have chosen $B_0=1.0$ nG and $n_B=-2.9$ at a magnetic coherence length scale $\lambda_B=1.0$ Mpc. In the plots shown in figure \eqref{fig:tb-sz-1}, on y-axis, we have considered absolute value of the $y-$parameter and on x-axis it is redshift $z$. We considered the absolute value of $y-$parameter since we chose the log-log scale, and any negative parameter value can not give a real value. So absolute value of the $y-$parameter is the best choice here. In the above plot, any kink in the plots represents the redshift at which the sign of the difference  $T_b-T_{\rm CMB}$ changes the sign. It is clear from the plot that decay of the dark matter can contribute to a large distortion of the CMB compared to the annihilation at low redshift. Inverse Compton scattering by high-energy electrons in galaxy clusters causes CMB-spectrum distortion, in which low-energy CMB photons acquire an average energy increase during the collision with high-energy cluster electrons. The redshift of the kink in case of decay depends on the chosen parameters. In references \citep{Minoda:2017iob, Chluba:2015lpa, Tashiro:2011hn}, authors have shown the impact of the magnetic fields on the absolute values of mean $y-$parameter. In this case the value of the parameter at $z\approx 10$ redshift for a quasi-scale-invariant PMF heating, leads to a maximum value of $y-$parameter: $|y|\sim 1.1\times 10^{-7}$ ($95\%$ confidence level). However in the case of BDM interaction$+B_0$ can contribute no more than $|y|\leq 10^{-9}$ \citep{Pandey:2020hfc}. Although measurements using a PIXIE-like experiment \citep{Kogut:2011djf} may achieve this sensitivity, the reionization and structure building processes alone cause a far higher distortion ($|y|\sim 10^{-7}-10^{-6}$) \citep{Hu:1993tc, Refregier:2000xz}. In a magneticum Pathfinder simulations \citep{WMAP:2010qai}, state of the art cosmological hydrodynamic simulations of large cosmological volume of $(896~\text{Mpc}~h^{-1})^3$ predicts the mean fluctuating Compton $y-$parameter of $1.18\times 10^{-6}$ which is well within the values obtained from the PLANCK data, i.e., $|y|\leq 2.2\times 10^{-6}$ \citep{Khatri:2015jxa, Dolag:2015dta}. Other best fitted value from the COBE/FIRAS measurement is $|y|< 5.5\times 10^{-5}$ \cite{Fixsen:1996nj}. Thus, regardless of the decay/annihilation channels of the DM particles, the maximum value of the mean $y-$parameter in the current scenario is well within the observational limits set by experimental data such as PLANCK, FIRAS, PIXIE, and others.
\subsection{Dark photons and the magnetic fields}\label{sec-3-DP-B0}
In this subsection, we determine the strength of magnetic field strength if we assume that these fields are originated from the decay/annihilation of the DM particles. In equation  \eqref{eq:baryon_temp-2}, in the absence of any background magnetic field, the second and third terms will not contribute to the thermal and ionization history of IGM. However, there may be a possible scenario when the decay/annihilation of the dark matter particles may contribute to the ordinary photons. In this case, if we compare the third and last term in equation \eqref{eq:baryon_temp-2}, for the decay of DM particles,
\begin{equation}
 	B_0(z)^2=\frac{f_{\rm eff}}{1+f_{\rm He}+x_e(z)}(1+z)^3\frac{f_{\chi,{\rm dec}}}{f_{\chi,{\rm in}}} \left(\frac{8 \pi\Omega_{\rm DM}c^2 \rho_c {\rm ln}(2)}{3 m H_0(z)}\right)\frac{1}{\tau}
 	\label{eq:B0-dec-chi-gamma-gamma}
\end{equation}
Similarly in the case of annihilation, we can obtain the formula
\begin{eqnarray}
    \scalemath{0.85}{B_0(z)^2=\frac{f_{\rm eff}}{1+f_{\rm He}+x_e(z)}(1+z)^6
    \frac{f_{\chi,{\rm ann}}^2}{f_{\chi,{\rm in}}}\left(\frac{8 \pi\Omega_{\rm DM}^2 c^2 \rho_c^2 {\rm ln}(2)}{3 m H_0(z)}\right)
 	\frac{\langle \sigma v\rangle }{m_\chi}}
	\label{eq:B0-ann-chi-gamma-gamma}
\end{eqnarray}
Therefore, we can see that the value of $B_0$ has a redshift depending on $z$, and it can be easily evaluated if we know the ionization fraction of the hydrogen atoms and the decay or annihilation channels.  
%
\begin{figure}
\centering
 	\includegraphics[width=0.82\linewidth, keepaspectratio]{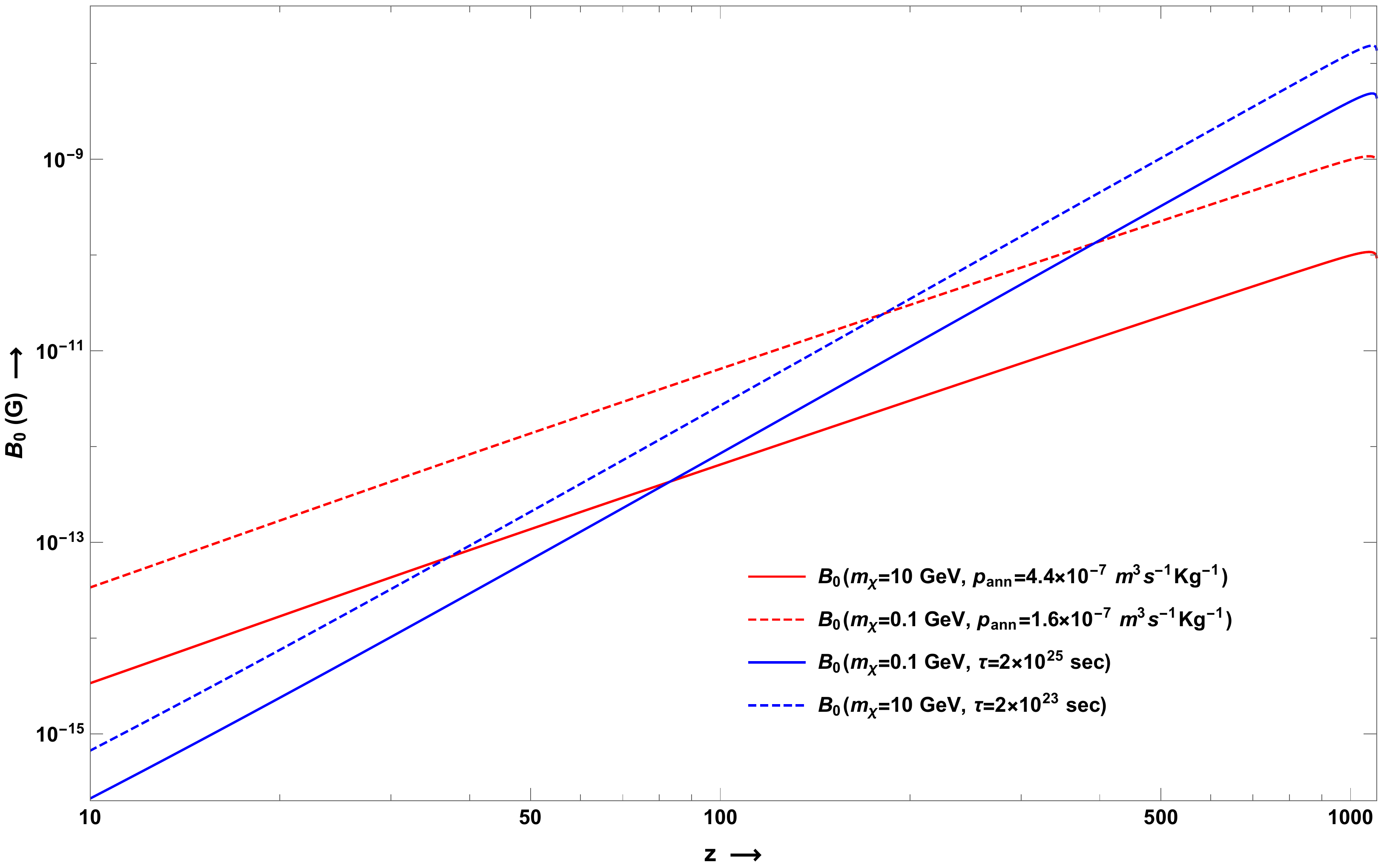}
 	\caption{\textbf{Strength of the magnetic fields:} Here we have assumed that there are no background magnetic fields and that $f_{\rm eff}=1.0$, $f_{\chi,{\rm ann}}=1.0$, $f_{\chi,{\rm dec}}=1.0$, $f_{\chi,{\rm in}}=1.0$, $n_B=-2.9$.}
 	\label{fig:B0-z-feb-2022}
\end{figure}
In the figure (\ref{fig:B0-z-feb-2022}), we have shown the variation of $B_0$ as a function of $z$ using equations \eqref{eq:B0-dec-chi-gamma-gamma} and (\ref{eq:B0-ann-chi-gamma-gamma}). For simplicity, we choose $f_{\rm eff}=1.0$, $f_{\chi,{\rm ann}}=1.0$, $f_{\chi,{\rm dec}}=1.0$, $f_{\chi,{\rm in}}=1.0$, $n_B=-2.9$. At redshift $z\approx 1000$, in case of annihilation, magnetic fields of strength is $B_0\approx  10^{-10} \mu G$ at a  length scale $\lambda_B \sim 10$ Kpc, which is equivalent to a present day comoving length scale of $\sim$ Mpc. These magnetic fields at redshift of 10, are equivalent to the field of the order of $\approx 10^{-15}$ G. Similarly for the decay channels, any $B_0\approx  10^{-11}$ G at recombination (equivalent to $\approx 10^{-16}$ G) at $z\approx 10$ at a length scale of $\sim 10$ Kpc (comoving length $\sim $ Mpc) can be obtained. Therefore, above calculated bound on the strength of the magnetic fields in the context of the DM decay/annihilation through various possible channels, could also affect the thermal and ionization history of IGM even in absence of any background magnetic fields. To understand the decay/annihilation channels and hence origin of the magnetic fields, we have briefly summarizes  processes involves below. 
 
As we have discussed above that the dark photons and the ordinary photons mix kinetically, the direct product of the groups corresponding to the two photons, i.e., $U(1)'\times U(1)$, eventually broken down to $U(1)_{\rm em}$ \cite{Berezhiani:1995am, Berezhiani:1995yi}. In this scenario, conversion of the dark photons to ordinary photons could explain the observed magnetic fields at a large scale \cite{Berezhiani:2013dea}. Before the recombination DM particles are strongly coupled to electrons and through them to photons. Therefore DM particles cannot participate in the structure formation. After the recombination epoch, when the ionization fraction drops below $10^{-4}$, the collision time becomes larger than the cosmological time. At this epoch, DM particles decouple from the baryon-photon fluid and do not follow the baryon bulk flow. At this epoch, the velocity of baryons and the DM particles are somewhat different. The drag force exerted by the DM particles on the electrons is given by $\vec{F}=e \Delta \vec{v} B_F$, where $\Delta \vec{v}$ is the relative velocity and $B_F=\sigma_{e\chi} n_\chi \omega_\chi/e$.  In the expression of the force $\vec{F}$, $n_\chi$ and $\omega_\chi$ are the number density and the mean energy density of the DM particles. The DM interaction with the protons is damped by the factor $(m_p/m_e)^2$. Therefore for a larger pressure on the electrons from the DM than CMB photons, cross-section $\sigma_{e\gamma}$ must be larger than $\sigma_{e\chi}$ at low momentum transfer. For the case when $m_\chi \approx 10~\text{keV}< m_e$, it has been shown that maximum values of $B_0$ could be of the order of $\sim 10^{-13}$ G in a galactic halos of size $10$ kpc \cite{Berezhiani:2013dea}.  However, for the case when $m_\chi >m_e$, the channel ($\chi\chi\rightarrow e^+ e^-$) could give a magnetic field of strength $\sim 10^{-15}$ G for $m_\chi \simeq 1$ MeV and $\sim 10^{-12}$ G for $m_\chi\simeq 1$ GeV at $\lambda_B\sim $ kpc scale. These channels at $\sim 1$ Mpc scale could also generate a magnetic field of the order of $\sim 10^{-14}$ G.
\section{Conclusion} 
\label{sec-4}
The IGM's thermal and ionization evolution is determined by the processes involved in influencing the dynamics of the baryons and the DM particles in IGM. Here we have included the following processes: heating from the magnetic fields, energy injection from DM decay/annihilation in the presence of non-standard interaction of baryon and the DM particles. To study the thermal and ionization history of the IGM, we have solved coupled equations of the baryon temperature $T_b$, DM temperature $T_d$, ionization fraction $x_e$ along with the relative velocity ${\rm v}$ and the magnetic energy evolution equations. We have also evolved the absolute value of the mean $y-$parameter of the thermal SZ effect for redshift $z$ to study the CMB anisotropy. The relative strength and nature of the magnetic field, on the one hand, and the BDM interaction cross-section, on the other, determine the nature of the IGM's thermal and ionization history when BDM interaction and magnetic fields are present. However, in the case when only DM decay or annihilation processes are considered, thermal and ionization history is solely determined by the lifetime of the DM particle or annihilation parameter $f_{\rm eff}\langle \sigma v \rangle/m_\chi$. It's worth noting that the energy deposited by DM decay or annihilation is considered to be instantaneous. However, this may not be the case, and the energy deposition's dependency on $z$ must be carefully studied. We also assume that only a fraction of the total DM particles interact with the baryons. The interaction is so that DM particles interact only with the charged baryons and have a non-zero small electric charge. We show that these processes significantly impact the baryon and DM temperature and the ionization fraction. This, in turn, leaves imprints through the tSZ effect on the CMB distortion at small scales. We show that the maximum value obtained for the absolute value of the mean $y-$parameter is well within the values obtained from the observational data of PLANCK, FIRAS, and PIXIE. We also calculate the magnetic field's strength, which is generated by the dark photon, a dark sector gauge field that is identical to the ordinary photon but is associated with the dark sector. We show that at a comoving scale of $\sim$ Mpc, the strength of the magnetic fields is of the order of $10^{-15}~\mu$G for annihilation channels and $10^{-16}~\mu$G for decay channels. In the current context, it is worthwhile to investigate the role of several other probable channels of DM particle decay/annihilation on the temperature and ionization history of the IGM. Especially at low redshift during the structure formation, the decay/annihilation processes could impact the gravitational collapse. 
\section*{Acknowledgments}
A.K.P. acknowledges the facilities at I.C.A.R.D., Department of Physics and Astrophysics, University of Delhi, Delhi, India. A.K.P. is financially supported by Dr. D.S. Kothari Post-Doctoral Fellowship provided by U.G.C., Govt. of India, under the Grant No. D.S.K.P.D.F. Ref. No. F.$4-2/2006$ (BSR)/PH/$18-19/0070$. Prof. T. R. Seshadri was also instrumental in motivating the author and providing valuable feedback and suggestions during the early stages of the project.
\section*{Data Availability Statement}
This manuscript has no associated external data [\textbf{Authors' comment:} This is a theoretical study. We have not used any experimental data]. The datasets generated during and/or analysed during the current study are available from the corresponding author on reasonable request.
\bibliographystyle{mnras}
\bibliography{mnrasref} 
\end{document}